\documentstyle[12pt,aaspp4,flushrt]{article}

\newcommand{\msb}{$M_{\odot}$~}
\newcommand{\ms}{$M_{\odot}$}

\begin{document}

\title{Inhomogeneous Chemical Evolution of the Galactic Halo: 
Abundance of $r$-process Elements}

\author{Claudia Travaglio$^{1,2}$, Daniele Galli$^3$, 
Andreas Burkert$^1$}

\bigskip

\affil{1. Max-Planck Institut f\"ur Astronomie, K\"onigstuhl, 17,
69117 Heidelberg, Germany}
\affil{2. Dipartimento di Astronomia e Scienza dello Spazio, Largo
E. Fermi 5, I-50125 Firenze, Italy}
\affil{3. Osservatorio Astrofisico di Arcetri, Largo E. Fermi 5, I-50125
Firenze, Italy}
 
\begin{abstract}

We present a stochastic model based on the Monte Carlo technique to
study the inhomogeneous chemical evolution of the Galactic halo.  In
particular, we consider the local enrichment and subsequent mixing of
the interstellar gas resulting from bursts of star formation, following
explicitly the fragmentation and coalescence of interstellar gas
clouds.  The model takes into account the mixing of halo gas through
cloud-cloud collisions and the delayed mixing of supernova ejecta into
the interstellar medium. The consequences of the infall of halo gas
onto the Galactic disk are also discussed. The mass spectrum of clouds,
the age-metallicity relation, and the G-dwarf distribution are
investigated at different times. We analyze in detail the predictions
of our model for the abundance of elements like Eu, Ba, and Sr in the
Galactic halo, following their production by $r$-process
nucleosynthesis for different assumptions on the supernova mass range.
Finally, we compare our results with spectroscopic data for the
chemical composition of metal-poor halo stars.

\end{abstract}


\section{Introduction}

The enrichment of the interstellar medium (hereafter ISM) in heavy
elements by successive generations of stars is a key issue in
understanding the chemical evolution of galaxies in general, and the
formation history and abundance distribution of stellar populations in
our Galaxy in particular.  From an observational point of view, very
metal poor stars provide important clues to reconstruct the
chemo-dynamical aspects of the first stages of halo evolution.
Therefore, a large body of spectroscopic abundance determinations has
been obtained for halo stars. In particular, observational studies
related to the heavy element abundances in the Galactic halo have shown
that stars of similar metallicity exhibit large abundance variations,
over about 2 dex in [element/Fe], for heavy elements like Eu, Ba and Sr
(see e.g.  Gilroy et al.~1988, Ryan \& Norris~1991, Gratton \&
Sneden~1994, McWilliam et al.~1995b, Ryan, Norris \& Beers~1996, and,
more recently, McWilliam~1998, Sneden et al.~1998, Burris et
al.~2000).  However, the actual nature of this scatter, and the
possible influence of various effects like different calibration
methods or data reduction procedures, is still unclear. If the observed
scatter is {\em intrinsic}, one is led to consider a scenario in which
the oldest halo stars were formed out of gas of strongly inhomogeneous
chemical composition.  For instance, the peculiar abundances determined
by Sneden et al.~(1994) and McWilliam et al.~(1995a) in the star CS
22892-052 (where $r$-process elements are enhanced over about 40 times
the solar value) provide support to this hypothesis. Similar studies of
objects in the Magellanic clouds (e.g. Cohen~1982, Da Costa~1991,
Olsewski et al.~1991) and dwarf galaxies (e.g. Pilyugin~1992, Kunth et
al.~1994) suggest that inhomogeneous chemical evolution is a common
phenomenon in nearby galaxies as well.

From a theoretical point of view, as first suggested by Truran~(1981),
the presence of $r$-process elements in low metallicity halo stars is
indicative of a prompt enrichment of the Galaxy in these elements,
possibly by early generations of massive stars.  Particular attention
to the Galactic evolution of elements produced by neutron-capture
nucleosynthesis was given by Mathews, Bazan, \& Cowan~(1992), Pagel \&
Tautvai\v{s}ien\.{e}~(1997), and more recently by Travaglio et
al.~(1999). These authors adopted the standard approach to Galactic
chemical evolution, assuming that stars form from a chemically
homogeneous medium at a continuous rate.  As stressed by Travaglio et
al.~(1999), this approach is able to reproduce spatially averaged
values of element abundances over the Galactic age, but a more
realistic model for the chemistry and dynamics of the gas is needed in
order to investigate the earliest phases of halo evolution.

Recently, several studies have attempted to follow the enrichment
history of the Galactic halo with special emphasis to the gas dynamical
processes occurring in the early Galaxy: Tsujimoto, Shigeyama, \&
Yoshii~(1999) provided an explanation for the spread of Eu observed in
the oldest halo stars in the context of a model of supernova-induced
star formation; Ikuta \& Arimoto~(1999) and by McWilliam \&
Searle~(1999) studied the metal enrichment of the Galactic halo with
the help of a stochastic model aimed at reproducing the observed Sr
abundances; Raiteri et al.~(1999) followed the Galactic evolution of Ba
by means of a hydrodynamical N-body/SPH code; finally, Argast et
al.~(2000) concentrated on the effects of local inhomogeneities in the
Galactic halo produced by individual supernova events, accounting in
this way for the observed scatter of some (but not all) elements
typically produced by Type~II supernovae.

In this paper we investigate whether incomplete mixing of the gas in
the Galactic halo can lead to local chemical inhomogeneities in the ISM
of the heavy elements, in particular Eu, Ba and Sr.  Our
chemo-dynamical evolution model for the Galactic halo is numerically
based on a Monte Carlo technique. Starting from a discrete distribution
of gas clouds, it follows star forming regions, stellar enrichment as
well as mixing processes, during the early time of the Galaxy. We also
investigate the consequences of infall of halo gas onto the disk. Since
accurate $r$-process yields cannot be obtained from current stellar
models, in order to study the composition of the halo gas in these
elements we use analytical calculations presented by Travaglio et
al.~(1999), summarized in Sect.~2.1.

Following the seminal paper by Searle \& Zinn~(1978), we represent the
Galactic halo as an ensemble of interacting gas clouds of different
chemical composition. At variance with other recent studies (e.g.
Tsujimoto et al.~1999) focused on the analysis of {\em local}
inhomogeneities within individual clouds, in our picture each cloud is
assumed to be chemically homogeneous, and able to host independent
episodes of star formation. Clouds are allowed to coalesce with each
other, and to fragment in smaller units, thus modifying the initial
mass distribution.  A similar scenario has been developed recently by
Smith~(1999) to study the problem of the formation of Galactic globular
cluster. According to Smith~(1999), if the proto-Galactic halo was
constituted of clouds of size $\sim 1$~kpc and mass in the range
$8\times 10^5$--$3\times 10^7$~\ms, evolving by coalescence, star
formation and fragmentation episodes, some properties of the Galactic
globular clusters system could be easily explained (e.g. their lower
cut-off in metallicity with respect to field stars, their chemical
homogeneity, etc.). In the present paper we do not investigate
explicitly the formation of globular clusters, but we assume instead
that stars formed in our clouds become part of the halo field star
population.

The paper is organized as follows: in Sect.~2 we summarize the main
characteristics of our chemo-dynamical model, the adopted stellar
yields and the IMF; in Sect.~3 we discuss the effects of episodic
infall of metal-deficient gas from the halo to the disk; in Sect.~4 we
present a series of test calculations as preliminary applications of
the Monte Carlo model presented here (i.e. a comparison with a ``simple
model'' of chemical evolution, the G-dwarf distribution, and the
evolution of the mass spectrum of coalesced interstellar clouds).  In
Sect.~5 we show the resulting age-metallicity relation, the abundances
of Eu, Ba, and Sr at the early epochs of the evolution of the Galaxy,
and we compare the model predictions with observations. Finally, in
Sect.~6 we briefly summarize the conclusions of our work, indicating
the constraints on the production of the $r$-process elements implied
by our analysis, and suggesting possible direction for future
investigations.

\section{A Stochastic Model for the Galactic Halo}

The main characteristics of the chemo-dynamical model presented in this
paper (summarized in Table~1) are a realistic treatment of chemical
inhomogeneities in the ISM due to incomplete mixing of stellar ejecta,
and the occurrence of discrete episodes of star formation localized in
time and space. We represent the halo gas by a ensemble of clouds of
different chemical composition, assumed uniform inside each cloud.  The
idea that interstellar clouds collide and grow by coalescence (process
in which clouds grow by accretion and interactions between other clouds
and become progressively more massive and condensed) was first
suggested by Hoyle (1953) and Oort (1954).  The cloud population
undergoes episodes of coalescence, fragmentation and star formation
bursts.  These processes are halted when the cloud reaches a critical
mass $M_{\rm cr}$ at which it becomes gravitationally unstable and
forms stars. In the present work, we consider the halo composed of
discrete gas clouds and we follow their evolution during the early
epoch of the Galaxy.  To follow each coalescence episode, we select one
cloud and choose randomly another cloud with a probability depending on
the mass of the two clouds and on their collision cross section
$\sigma_{ij}$. This selection of pairs of clouds at a certain timestep
continues until all the clouds are examined, and occurs with a mixing
frequency $f_{\rm m}$.  The collision probability between two clouds of mass
$M_i$ and $M_j$ is
defined as
\begin{equation}
P_{ij}=\frac{\sigma_{ij}} {\sigma_{\rm max}} g(M_i,M_j),
\end{equation}
where
\begin{equation}
g(M_i,M_j)=\frac{M_i M_j} {M_{\rm max}^2}
\end{equation}
is a cut-off function adopted in order to minimize the probability of
uninfluential collisions between a low-mass and a high-mass cloud. 
The probability $P_{ij}$ is normalized to the values of the mass $M_{\rm
max}$ and cross section $\sigma_{\rm max}$ of the most massive cloud at
each timestep.

\begin{table}[t]
\begin{center}
TABLE 1\\
{\sc Parameters of the Standard Model}\\
\vspace{0.5em}
\begin{tabular}{ll}
\hline
initial number of clouds $N$          & $10^4$ \\
total initial mass $M_{\rm halo}$     & $5\times 10^{10}$~\ms \\
initial mass range of clouds          & $10^3$--$10^7$~\ms \\
burst frequency $f_{\rm b}$           & $5\times 10^{-7}$~yr$^{-1}$ \\
mixing frequency $f_{\rm m}$          & $2.5 \times 10^{-7}$~yr$^{-1}$\\
star formation timescale $t_0$        & $2\times 10^7$ yr\\
critical cloud mass $M_{\rm cr}$      & $10^4$~\ms \\
star formation efficiency $\eta$      & 0.03  \\
IMF                                   & Salpeter 0.1--120~\ms \\
mass range of SNII                    & 8--120~\ms \\
$e$-folding time of disk accretion    & $10^9$~yr \\
\hline
\end{tabular}
\end{center}
\end{table}

The clouds are assumed to be spherical with constant density, and 
the effective collision cross section $\sigma_{ij}$ is proportional to
the geometrical cross section
\begin{equation}
\sigma_{ij} \propto R_i^2 + R_j^2, 
\end{equation}
where $R_i$, $R_j$ are the radii 
the clouds $i$ and $j$. Observations suggest the existence of a
mass-radius relation of present-day molecular clouds (e.g. Blitz~1993) in
the Galactic disk
\begin{equation}
M \propto R^2 ,
\end{equation}
resulting in a collision cross section linearly proportional to sum
of the masses of the clouds
\begin{equation}
\sigma_{ij} \propto M_i + M_j, 
\end{equation}  
which is the expression for the collision cross section adopted in this
paper. The new cloud formed in this coalescence episode has a mass
equal to the sum of the masses of the two parent clouds, chemical
composition equal to their mass-averaged chemical compositions, and
zero age. Each new cloud formed at a certain timestep is not allowed to
participate again in coalescence episodes at the same timestep.  As a
result of the process of coalescence, more massive clouds are produced
at each timestep until they reach a critical value $M_{\rm
cr}=10^4$~\msb at which they are allowed to form dense molecular cores
and give birth to clusters of stars.  We assume a star formation
efficiency $\eta=0.03$ in each cloud, i.e. each burst converts 3\% of
the mass of the parent cloud $M$ into stars.

Following a stellar burst, a cloud breaks up into a distribution of
smaller clouds, due to the energetic processes that accompany star
formation. Numerical calculations of the fragmentation process favour a
small number of fragments ($\sim$ 2 to 5) per fragmentation stage (see
e.g. Scalo~1985). We assume therefore that the original cloud produces
a random number of fragments from 1 (no fragmentation) to 5 fragments
(maximum fragmentation) characterized by the same chemical composition
of the parent cloud. The age of the newly created cloud fragments is
reset to zero.  To estimate the timestep for the fragmentation episode
we assume that, since the free-fall time in a molecular cloud is $\sim
10^6$~yr, stars are not allowed to form on a timescale smaller than
$\sim 10^6$~yr.  Observations and theory suggest that low- and
intermediate-mass stars form within dense cores of molecular clouds on
a timescale of $\sim$~10$^7$~yr (e.g. Shu et al.~1993). Hence we assume
that the probability of a star formation burst (whenever the cloud's
mass is higher than $10^4$~\ms) is a Gaussian function of the cloud's
age $t$
\begin{equation} 
P(t) = C \exp\left[-\left(\frac{t-t_0}{t_0}\right)^2\right], 
\end{equation}
with $t_0 = 2\times 10^7$~ yr.  As shown in Table~1, the star formation
rate (hereafter SFR) in our model occurs with a frequency $f_{\rm
b}=5\times 10^{-7}$~yr$^{-1}$.

The SFR, under our assumptions, depends linearly on the mass of the
most massive clouds.  An order-of-magnitude value of the SFR can be
estimated from the ratio between the mass of the gas able to form stars
and the timescale of evolution of the Galactic halo.  If one assumes
that the baryonic mass of the Galactic halo is $\sim$
5$\times$10$^{10}$ \ms, and the efficiency of star formation is a few
percent, the mass converted in stars in the Galactic halo is a few
$10^9$~\msb. For the lower limit of the timescale of evolution of the
halo we can use the free-fall time $\tau_{\rm ff}$.  Following the
pioneering work by Eggen, Lynden-Bell \& Sandage~(1962), and the work
by Fall \& Rees~(1985), assuming a pre-existing dark halo of radius
$R$, which would account for the observed flat rotation curve of
rotational velocity $v_{\rm c}\simeq$ 220 km s$^{-1}$, then $\tau_{\rm
ff}$ is given by
\begin{equation}
\tau_{\rm ff}=2.8\times10^8\left({R\over 50~{\rm kpc}}\right)~{\rm yr}.
\end{equation}
For the star formation timescale in
the halo we can use the range between $\tau_{\rm ff}$ and $\sim
4\tau_{\rm ff}$. Therefore, the expected SFR, given by the ratio
(mass converted into stars)/$\tau_{\rm ff}$, is of order 
a few \ms~yr$^{-1}$.

We have explored different ranges for the initial mass of the clouds
(the initial distribution of the clouds in this mass range being always
chosen randomly), finding that a range of $10^3$--$10^5$~\msb leads to
an average SFR which is too low (with an average value at
0.01~\ms~yr$^{-1}$), as shown in Fig.~1 (upper panel). An average SFR
of $\sim$ 1.5~\ms~yr$^{-1}$ can be obtained starting with an initial
mass range between $10^3$--$10^7$~\msb as shown in Fig.~1 (lower
panel). We also found that SFR is linearly proportional to the
efficiency $\eta$ and to the ratio $f_{\rm m}/f_{\rm b}$. Then, to
obtain a SFR of order a few \msb yr$^{-1}$ with the lower mass range of
clouds (10$^3$--10$^5$ \ms) we need extreme values of $\eta$ ($\sim 1$)
or $f_{\rm m}/f_{\rm b}$ ($\sim 10^3$ times higher than our
standard values reported in Table~1).

As suggested earlier by Talbot \& Arnett~(1973) and Edmunds~(1975), the
timescale for the mixing of gas in the Galaxy, associated with random
motions of the clouds, can be longer than the star formation time
scale. In fact, the extent of dilution which occurs before stars form
depends upon the interval between the supernova events and the next
star formation event, and it depends upon whether mixing will occur
throughout this time interval.  For the simulations presented here we
have explored different choices for the frequency of coalescence with
respect to the burst frequency (as will be discussed below), concluding
that $f_{\rm b}\sim 2 f_{\rm m}$ represents the best value.

In this work we start with $N=10^4$ clouds, randomly distrubuted in the
mass range $10^3$--$10^7$~$M_\odot$, and with a total mass $M_{\rm
halo}=5\times 10^{10}$~\ms.  The mass spectrum of the stars formed in
the burst is described by a Salpeter stellar initial mass function
(hereafter IMF) \begin{equation} \frac{d{\cal N}}{dm}(m)=Am^{-2.35},
\end{equation} where $m$ is the mass of the star in solar masses and
$A$ is a normalization constant.  If the range of stellar masses
extends from 0.1~$M_\odot$ to 120~$M_\odot$, then $A$ is be determined
from the condition
\begin{equation}
\int_{0.1}^{120}m\frac{d{\cal N}}{dm}(m)dm=\eta M,
\end{equation}
giving
\begin{equation}
A\simeq
5.1\left(\frac{\eta}{0.03}\right)\left(\frac{M}{10^3~M_\odot}\right).
\end{equation}
The sensitivity of the results from the adopted IMF will be discussed
below.

\subsection{Adopted Stellar Yields}

We concentrate in this paper on the early enrichment of the halo gas
($t \le 1$~Gyr) in Fe and heavy elements, in which the dominant role is
played by Type~II SNe (SNII). Therefore we include only the contribution of
SNII without considering Type Ia supernovae. The Fe yields of SNII are
taken from Woosley \& Weaver~(1995, hereafter WW95), for the mass range
12--40~\msb (model ``A'', with $Z=10^{-4}Z_\odot$ and $Z=0$), and the
delayed contribution to the chemical enrichment of the halo gas from SNe
of different masses is also taken into account.

As no calculation for progenitors more massive than 40~$M_\odot$ is
available in WW95, for $m > 40$~$M_\odot$ we have explored various
stellar nucleosynthesis calculations recently published for massive
stars (e.g. Thielemann, Nomoto \& Hashimoto~1996; Chieffi, Limongi \&
Straniero~1998; Portinari, Chiosi, \& Bressan~1998; Woosley, Langer \&
Weaver~1995; Maeder~1992).  The results shown in this paper are taken
from Woosley, Langer \& Weaver~(1995) for a star of 60 $M_\odot$ and
solar metallicity (model 7K) in order to have a self-consistent data
set for a wide range of stellar masses.

The production sites of $r$-process elements still need to be
unambiguously identified, despite the large number of recent studies
(see e.g. the hydrodynamic simulations by Wheeler, Cowan \&
Hillebrandt~1998, and Freiburghaus et al.~1999), and quantitative
estimates of the $r$-process yields are still not available.  Tsujimoto
et al.~(1999), Ikuta \& Arimoto~(1999), and McWilliam \& Searle~(1999)
deduced empirically the the $r$-process yields from the available
observational data of the most metal-poor stars, whereas Travaglio et
al.~(1999) presented analytical estimates of the $r$-process yields
independent of observations, treating the $r$-process as a {\it
primary} process originating in SNII. In this approach, the so called
$r$-residuals are derived after subtracting from the solar abundances
the predicted $s$-fractions at $t=t_\odot$ (see Travaglio et. al.~1999
for details).  At lower metallicities the $r$-process yield is assumed
to be proportional to the oxygen yield of SNII.

In the present work, to follow the evolution of Ba and Eu, we adopt
the $r$-process calculations for different metallicities presented by 
Travaglio et al.~(1999). Moreover, as will be discussed in more detail,
the spectroscopic observations of [$r$/Fe] as a function of [Fe/H] show
a decline of [$r$/Fe] at lower metallicities, that can be naturally
explained by a time delay between the O-rich (partly Fe-rich) material
ejected by the more massive SNe ($m\ge 15$~\ms), and the $r$-process
material ejected by the lower mass SNe (8--10~\msb with zero Fe
production).  This delay may reflect the differences in stellar lifetime,
but can be amplified by non-instantaneous mixing processes in the ISM.
Finally, we have introduced new assumptions to quantify the production
of Sr from massive stars, discussed in more detail in Sect.~5.3. In
Table~2 are summarized the averaged stellar yields calculated for this
work for Eu and Ba (in the mass range 8--10~\ms), and for Sr (in the
mass range 15--25~\ms).

\begin{table}[t]
\begin{center}
TABLE 2\\
{\sc Yields and Abundances of Eu, Ba, and Sr Adopted in This Work}\\
\vspace{0.5em}
\begin{tabular}{llll}
\hline
 & Mass range & $\langle$Yield$\rangle$ & Solar mass fraction \\
 & (\ms) & (\ms) & (Anders \& Grevesse~1989) \\ 
\hline
Eu & 8--10  & $6.4 \times 10^{-7}$ & $3.74 \times 10^{-10}$ \\
Ba & 8--10  & $5.7 \times 10^{-6}$ & $1.56 \times 10^{-8}$ \\
Sr & 15--25 & $3.5 \times 10^{-6}$ & $5.16 \times 10^{-8}$ \\
\hline
\end{tabular}
\end{center}
\end{table}

\subsection{The Stellar Initial Mass Function}

Several observational studies have been carried out in order to
establish space or time variations of the IMF in our Galaxy and their
dependence on parameters such as the gas metallicity.  The regions that
have been studied with direct star counts so far include the local
field star population in our Galaxy and many star clusters of all ages
and metallicities in both our Galaxy and Magellanic Clouds.  The IMF
derived for field stars in the solar neighbourhood exhibits a power-law
decline with mass above $\sim 1$~\msb roughly consistent with the
original Salpeter~(1955) law (Miller \& Scalo~1979, hereafter MS; for a
more recent review, see e.g. Larson~1998). However, below this value,
the IMF of field stars flattens, showing a possible broad peak at a few
tenths of a solar mass in the number of stars per unit logarithmic mass
interval.  However, the IMF at these masses is uncertain because it
depends, among other things, on the poorly constrained evolutionary
history of the Galaxy.

In this paper we adopt a Salpeter IMF but we 
explore the sensitivity of the model results (in particular the
age-metallicity relation and the G-dwarf distribution) to different
choices of the IMF.  For instance, adopting the analytical expression
of the IMF given by MS, the role of low-mass and high-mass stars is
respectively overestimated and underestimated compared with the
Salpeter IMF. The ratio between the number of massive stars in the mass
range 8--120 \msb calculated with the MS IMF and the Salpeter IMF,
for the same initial cloud mass, is
\begin{equation}
\frac{({\cal N}_{8-120})_{\rm MS}}{({\cal N}_{8-120})_{\rm Salpeter}}
\simeq 0.03.
\end{equation}
Since the only Fe stellar sources for the first Gyr of the evolution of
the Galaxy are SNII in this mass range, with a MS IMF the metallicity
versus time covers a lower range of values.  In particular, with this
model we obtained [Fe/H] values that converged at 1 Gyr at [Fe/H] $\simeq
-3$, instead of [Fe/H] $\simeq -1.5$ as in the case of a Salpeter IMF.
In addition, with a MS IMF (underlying all the other assumptions of the Monte
Carlo model presented here), in order to reach a metallicity [Fe/H]
$\simeq -1.5$, the efficiency of star formation has to be increased by
a large factor, from 3\% with the Salpeter IMF up to 60\% with the MS IMF.

\section{Infall}

As first stressed by Larson~(1972) and by Hartwick~(1976), gas infall
from the Galactic halo to the disk guarantees additional material
available above the amount that has been consumed, sustaining the disk
star formation for longer times.  Theoretical constraints on the
collapse of the Galactic halo can be obtained by comparing its
evolutionary and dynamical timescales.  Hydrodynamical simulations by
Burkert, Truran \& Hensler~(1992), and by Truran \& Burkert~(1993)
showed that if the proto-Galactic gas was initially shock heated to the
virial temperature $T_{\rm vir}\sim 2\times 10^6$~K (or if early
energetic star formation had acted to heat the halo to its virial
temperature), the cooling timescale would be $\sim 3\times 10^9$~yr
(see also Boehringer \& Hensler~1989), which is larger than the
free-fall timescale.  This result indicates that energetic heating and
cooling processes might play an important role in regulating the
early stages of formation and collapse of the Galactic halo. These
theoretical arguments suggest a halo formation and collapse timescale
of the order of $\sim 10^9$ yr. The available abundance data for field
and globular cluster stars in our Galaxy, when interpreted in the
context of the current knowledge of nucleosynthesis theory, allow
interesting timescale constraints to be imposed on the earliest stages
of Galactic evolution. The implied timescale of star formation and
supernova nucleosynthesis activity in the halo of our Galaxy appears to
exceed the best estimates of several $10^8$ yr, available from
dynamical arguments. It seems increasingly clear that a rather complex
sequence of events as a result of hierarchical mergers defined the
formation and early evolution of the Galactic halo and disk
components.  Future abundance studies, particularly for globular
cluster and field stars, may reasonably be expected to provide critical
clues to the nature of this evolution.  In the model presented here we
assume that the gas mass in the halo decreases exponentially with an
e-folding time $\tau=10^9$~yr.  Following these assumptions, at 1 Gyr,
when the residual gas mass is reduced by $\sim 40$\%, the metallicity
is [Fe/H] $\simeq -2$.

\section{Test Calculations}

In this Section we present a series of test calculations as preliminary
applications of the Monte Carlo model. In particular, the current gas
fraction, the distribution of long-lived stars in metallicity (G-dwarf
problem), as well as the evolution of the mass spectrum of interstellar
clouds due to coalescence episodes, are used to determine the values of
the free parameters of the model.

\subsection{Test 1: Comparison with the ``Simple Model'' of Chemical
Evolution}

As a first test of our model, we compare the distribution of Fe
abundance versus the gas fraction resulting from our Monte Carlo
simulations and from a ``closed box'' model of Galactic evolution. The
``closed box'' model, as introduced by Tinsley~(1974), is characterized
by a SFR $\psi(t)$, total mass $M_t$ and gas mass $M_g(t)$ in the form
of gas, in the instantaneous recycling approximation (IRA). The system
is closed, with $M_g(0) = M_t$. The evolution of gas mass and metal
content is regulated by the following expressions
\begin{equation}
\label{ms1}
\frac{d M_g}{dt} = -(1-R)\psi ,
\end{equation}
and
\begin{equation}
\label{ms2}
\frac{d} {dt}(M_gZ) = -Z(1-R)\psi + P_Z\psi ,
\end{equation}
where $R$ is the returned fraction of gas defined as
\begin{equation}
\label{ms3}
R = \int^{m_{\rm max}}_{m_1} [1-d(m)]\phi(m)dm ,
\end{equation}
and $P_Z$ is the primary metal production factor defined as
\begin{equation}
\label{ms4}
P_Z = \int^{m_{\rm max}}_{m_1} [q_{\rm c}(m) - d(m)]\psi(m)dm . 
\end{equation}
In these expressions, $\phi(m)$ is the IMF, $d(m)$ defines the mass
fraction which remains as a stellar remnant upon the death of the star
(i.e. all the matter external to $d(m)$ is ejected in the ISM), and
$q_{\rm c}(m)$ corresponds to the fraction (in mass) of $^4$He converted into
C, O, and heavier species.  The lower limit of these integrals $m_1$ is
the turn-off mass, taken as $m_1=0.8$~\msb for a Galactic age of 15 Gyr.
In our simulations we are interested only in the first Gyr of the
evolution of the Galaxy, therefore we take $m_1=2$~\ms.  The upper
limit is $m_{\rm max}=120$~\ms. 
A straightforward integration of eq.~(12)
and (13) gives, independently on the SFR, 
\begin{equation}
\label{ms5}
Z=-\frac{P_Z}{1-R}\ln\left(\frac{M_g}{M_t}\right).
\end{equation}
Inserting in eq.~(12)--(15) the values of the stellar
parameters adopted in our Monte Carlo model we obtain $R=0.71$,
$P_Z=7.9\times 10^{-4}$.  In Fig.~2 we compare the predictions of
eq.~(16), shown by the continuous line with the results of the Monte
Carlo model (open circles). For the latter, the average Fe abundance was
computed by summing the Fe mass in all clouds and dividing by the total
mass of gas at various timesteps (infall was turned off in this
simulation).  The analytical results approximate very well the
abundance of Fe computed numerically with the Monte Carlo model. In
fact the IRA approximation, implicitly adopted in the equations of the
``simple model'', is satisfied to a very high degree in our Monte Carlo
simulation, since the time evolution of the system ($\sim 10^9$~yr) is
much larger than the lifetime of the lower mass SNII stars ($\sim
10^7$~yr). However, the analytical calculations can only 
predict the average Fe composition at each timestep, while our 
Monte Carlo model can follow the evolution of metallicity distribution
of the halo clouds.

\subsection{Test 2: the G-dwarf Distribution}

The stars of spectral type G and luminosity corresponding to the dwarf
class are low-mass stars ($m \simeq 0.8$~\ms) evolving on a timescale
of about 15 Gyr, comparable to the estimated age of the Galaxy.
Therefore, they represent a sample that has never been affected by
stellar evolution, accumulating since the first episodes of low-mass
star formation.  The paucity of metal-poor stars in the solar
neighbourhood relative to the predictions of simple models of chemical
evolution was noted by van den Berg~(1962) and discussed further by
Schmidt~(1963) (see also Larson~1998). The G-dwarf problem has recently
been found also in other galaxies, including ellipticals (Bressan,
Chiosi \& Fagotto~1994; Vazdekis et al.~1996; Worthey, Dorman \&
Jones~1996).  To obtain the G-dwarf distribution we proceed in the
following way. Given the star formation rate $\psi(t)$ at a certain
time $t$ and defining as $\varphi_G$ the fraction of G-dwarfs in a
single stellar generation, the number ${\cal N}(z_1,z_2)$ of G-dwarfs
with $z_1 \le$ [Fe/H] $\le z_2$ is obtained as
\begin{equation}
{\cal N}(z_1,z_2)=\varphi_{\rm G}\int_{t(z_1)}^{t(z_2)}\psi(t)dt,
\end{equation}
where $t(z)$ is the time at which [Fe/H]$=z$.

The halo metallicity distribution function provides direct information
about the initial stages of galaxy formation as it is sensitive to the
bulk chemical properties of the interstellar gas from which the
earliest generation of stars were born. A compilation of the relative
numbers of low metallicity stars in the halo enables comparisons with
alternative models of Galactic chemical evolution, and can be used to
place constraints on the primordial rate of SNII, the SFR, and the time
scale for the element enrichment in the early Galaxy.  Early attempts
to address this problem were based on the metallicity distribution of
globular clusters determined by Hartwick~(1976) (see also Bond~1981).
Recent extensive observational campaigns by by e.g. Laird, Carney \&
Latham~(1988), Ryan \& Norris~(1991), Beers, Preston \&
Shectman~(1992), Carney et al.~(1994), and Beers et al.~(1998) have
largely extended and updated the original data set.  In particular,
Beers and coworkers have identified several hundreds of stars belonging
to the halo (and possibly thick disk) population of the Galaxy, and
combined these data with other samples of extremely metal-deficient
stars (Ryan \& Norris~1991; Beers et al.~1992; Carney et al.~1994) to
form a large database of low metallicity stars.  In this work, we take
as a reference the sample of low metallicity, high galactic latitude
halo stars obtained by Beers et al.~(1998, see their Fig.~2).

A comparison between the predictions of the Monte Carlo model for four
different values of the mixing frequency $f_{\rm m}$ and the
observational results by Beers et al.~(1998) is shown in Fig.~3. The
errorbars for the observational data indicate the $\sqrt{N_{stars}}$
noise associated with each bin. We also show in the same Figure the
G-dwarf distribution obtained with the simple model of chemical
evolution in the IRA approximation.  The best fit to the observations
is for the case $f_{\rm m}\simeq 2 f_{\rm b}$ which is assumed
hereafter as the {\it standard} for our simulations.  For larger values
of this ratio, the metallicity distribution predicted by our model
shows a substantial excess of stars below [Fe/H] $\simeq -3$ as
compared with observations.  In fact, lower metallicity clouds form
through coalescence episodes with clouds of primordial abundance
(higher metallicity clouds result from internal bursts of star
formation), and increasing $f_{\rm b}$ with respect to $f_{\rm m}$
results in the creation of more low-metallicity clouds (and
consequently low metallicity stars). At [Fe/H] $\simeq -2.5$ a
contribution from thick disk stars can affect to the model
predictions.

\subsection{Test 3: Coalescence and the Evolution of the Cloud Mass Spectrum}

In this Section we consider the evolution of the mass spectrum for
interstellar clouds as a result of coalescence episode alone, without
considering star formation and cloud fragmentation, in order to test
our model against well-known analytical results for the evolution of
the cloud mass spectrum under these circumstances (see e.g. Silk \&
Takahashi~1979).  For instance, Hayashi \& Nakagawa~(1965) found that
with a collision cross section scaling as $\sigma(M_1,M_2)\propto
M_1^{\lambda/2} M_2^{\lambda/2}$ the solution of the coagulation
equation must approach asymptotically the similarity solution 
defined by
\begin{equation}
t^{\frac{4+\lambda}{2-\lambda}}N(M,t) = f(t^{-\frac{2}{2-\lambda}}M).
\end{equation}
Since $\lambda=1$ in our standard case (see Sect.~2 for discussion),
we expect the mass spectrum to evolve toward a limiting form 
\begin{equation}
t^5 N(M,t) = f(t^{-2}M).
\end{equation}

We start our simulations with $N=10^4$ clouds with the same mass
$M=10^3$~\ms, and we follow the evolution of the mass spectrum for a
certain number of timesteps $N_{ts}$, until the cloud mass spectrum
reaches a quasi equilibrium distribution.  As mentioned above, the
coalescence episodes are regulated by selecting one cloud and choosing
randomly another cloud, with a probability that depends on the mass of
the two clouds and on their collision cross section. This selection of
pairs of clouds at a certain timestep continues until all clouds are
examined and have had the possibility to be involved in a coalescence
episode at a certain timestep.  We show in Fig.~4 the resulting
values of $t^5N(M,t)$ vs.  $t^{-2}M$ at different timesteps.  The shape
of the mass spectrum is clearly converging, after the first few
timesteps, toward a shape-invariant form that depends only on the
assumed dependence of the collision cross section on the masses of the
clouds.

\section{Results for the Chemical Evolution of the Galactic Halo}

In this Section we present the results for the age-metallicity relation
and for the abundances of Eu, Ba and Sr in the first Gyr of the Galaxy,
by considering their $r$-process contributions only. Being essentially a
pure $r$-process element, Eu is clearly a diagnostic for the chemical
enrichment of the gas by $r$-process products. For Ba and Sr one has to
disentangle the $s$- and $r$-process contribution weighted on the
relative stellar sources at different times during the evolution of the
Galaxy. On the other hand, Ba and Sr are observable at [Fe/H] $\simeq
-4$, whereas Eu is observable only to [Fe/H]$\simeq -3$, hence
predictions and models for the evolution in the halo gas of Ba and Sr
are better testable than Eu.

In the following, we show the results obtained with our Monte Carlo
model for Fe, for the Fe-group elements Mn and Co, and for the
heavier elements Eu, Ba and Sr vs. [Fe/H] in the halo gas, discussing
also the effect of variations of the standard values of the free
parameters of our model.

\subsection{Age-Metallicity Relation}

From an observational perspective, the age-metallicity relation
(hereafter AMR) for stars in the solar neighbourhood (mostly disk
stars) was first determined by Twarog~(1980), and later reanalyzed by
Meusinger et al.~(1991). A more recent study for the disk AMR, based on
a new sample of nearby F and G dwarfs, was performed by Edvardsson et
al.~(1993) using high resolution spectra to determine the surface
chemical composition of these stars.  Their results show clearly that
the concept of a well defined tight age-metallicity relation is
unfounded:  the slope of [Fe/H] vs. age over the lifetime of the
Galactic disk is flat with a large scatter in metallicity at all ages.
Unfortunately, a direct test of the age-metallicity 
relation for halo stars is not possible, since accurate 
age determinations are not available for this stellar population. 

In Fig.~5 we show the results for the distribution of [Fe/H] versus
time in the Galactic halo for different choices of the ratio $f_{\rm
m}/f_{\rm b}$, obtained with the values of the parameters listed in
Table~1 and described in Sect.~2.  It is clear that a well defined AMR
has to be replaced by a statistical relation.  From Fig.~5 one can
notice that after about $10^7$~yr the spread in [Fe/H] increases,
covering a range $-6<$ [Fe/H] $<-2$. This is due to coalescence
episodes between clouds that experienced at least one burst of star
formation and clouds that never experienced star formation episodes.
After a certain time which is of the order of $\sim 10^8$~yr, the halo
gas starts to homogenize its chemical composition, and the spread in
[Fe/H] is considerably reduced, converging to [Fe/H] $\simeq -2$.  The
time required for this ``homogenization'' depends clearly on the ratio
$f_{\rm m}/f_{\rm b}$, ranging from $\simeq 10^8$~yr for the case
$f_{\rm m}=f_{\rm b}$ to a few times $10^8$~yr for the cases with
higher $f_{\rm m}$.  This fact might have relevant consequences on the
distribution of [Eu/Fe], [Ba/Fe], and [Sr/Fe] vs.  [Fe/H] as will be
discussed in the following Section.

The distribution of [Fe/H] in the clouds is also influenced by the
number of SNe that explode during each burst and by the mass
distribution of clouds at each timestep. The upper limit on the cloud
mass range has considerable consequences for the number of high-mass
stars (50 to 100 $M_\odot$) formed during a star burst, since more
massive clouds have more chance to form stars in the high-mass tail of
the IMF (for a given efficiency $\eta$). This, in turns, affects the
evolution of elements produced by massive SNe (e.g.  oxygen), but has
no consequences on the elements considered in this paper. In a
forthcoming paper we will consider quantitatively the processes
limiting the growth of clouds by coalescence (tidal disruption, etc.)
and the related aspects of massive stars nucleosynthesis.

With the adopted Salpeter
IMF one can easily derive the number ${\cal N}_{\rm SN}$ of SNe that
contribute to the minimal enrichment of [Fe/H] in a cloud
\begin{equation}
{\cal N}_{\rm SN}=\int_{8}^{120}\frac{d{\cal N}}{dm}(m)dm
\simeq 0.13\left(\frac{\eta}{0.03}\right)
\left(\frac{M}{10^3~M_\odot}\right).
\end{equation}
If $M = 10^4$~\msb (corresponding to the lowest mass allowed to form
stars) and $\eta=0.03$, one obtains ${\cal N}_{\rm SN}\simeq 2$.  Ryan,
Norris \& Bessel (1991) argued that the ejecta of a single SN of 25
$M_\odot$ exploding in a $10^6$~$M_\odot$ cloud is sufficient to enrich
a cloud to [Fe/H] $= -3.8$. We find that the first enrichment episode
by the most massive SNe result in [Fe/H] $\simeq -5$ after a few
timesteps (at the first timesteps also the most massive SNe did not
have enough time to contribute to the enrichment of the gas). At $t
\sim 10^7$~yr the average metallicity has reached the value of [Fe/H] $\simeq
-3$, due to the important contribution to Fe enrichment from $\sim
20$~\msb SNe. Ryan et al.~(1991) maintained that the value of [Fe/H] $=
-3.8$ sets a lower limit on the metallicity of the second generation of
stars.  Under the assumptions adopted for this work, as one can see in
Fig.~5, Pop.~II stars can reach lower values of [Fe/H] even at later
times (up to [Fe/H] $\simeq -6$), due to efficiency of coalescence
episodes between clouds with different chemical composition.

The highest values of metallicity reached with these simulations (shown
in Fig.~5) at $10^9$~yr are [Fe/H] $\simeq -1.8$. It is possible to
reach higher [Fe/H] values by increasing the star formation efficiency
above the standard value $\eta=0.03$. In Fig.~5 is also shown for
comparison (with a solid line) the age-metallicity relation obtained
with the chemical evolution model presented by Travaglio et al.~(1999).
In Fig.~6 we show the age-metalliticy distribution for the case
$f_{\rm m}=2f_{\rm b}$ with $\eta=0.3$.  For this case [Fe/H] reaches
final values $\sim 1$~dex higher with respect to the case shown in
Fig.~5 ({\it upper right panel}).  Notice however that with a higher
star formation efficiency the spread covered by the [Fe/H] distribution
is higher, with significant consequences for the [Eu/Fe] and
[Ba/Fe] distribution vs. [Fe/H], as will be discussed in the following
Section. 

As a final test of our model, we have considered the evolution of an
ensemble of non-interacting interstellar clouds, switching off both
coalescence and fragmentation. In this case, each cloud evolves as a
closed box, continuosly enriched by episodes of star formation.  The
resulting age-metallicity relation shows no spread, demonstrating that
the scatter shown in Fig.~5 is a result of mixing among clouds of
different composition.

\subsection{Iron-group elements: Co and Mn}

As another test of the chemical evolution model described in this
paper, we consider the gas enrichment (limited to the halo phase) of
two of the most significant Fe-groups elements, Co and Mn. Although
both the observed abundances and the theoretical yields of Fe-group
elements are affected by large uncertainties, their study is of the
highest importance to test chemical evolution predictions and to
provide constraints on nucleosynthesis theory.

A remarkable result of recent observations of extremely metal-poor stars was
the discovery of two different metallicity dependences of Fe-group
element abundances (McWilliam et al.~1995b; Ryan, Norris, \&
Beers~1996): in particular, the two elements immediately below Fe, i.e.
Cr and Mn, drop to high overdeficiencies at [Fe/H]$<-2.5$ (in terms of
[X$_i$/Fe]), whereas elements above Fe, i.e. Co and possibly Ni, are
seen to rise at higher values. However, one should keep in mind that
the observational data can be affected by various uncertainties:  for
example, lines around $\sim 4000$~\AA can be blended with nearby strong
Fe I lines, leading to different estimates of the Mn or Co abundance in
the same objects.  Also, the solar photospheric and meteoritic
abundances of Mn differ by $\sim 0.14$ dex.

Both Mn and Co are synthesized mainly during explosive Si burning in
SNe (products of $^{55}$Co and $^{59}$Cu decay, respectively).
Therefore their yields are strongly affected by various factors like,
e.g., the position of the mass cut between the neutron star and the
ejecta, the total explosion energy and entropy, the delay time between
the collapse and the explosion, etc.. Since we concentrate on the early
stages of Galaxy evolution, we also need to take into account the
dependence of Co and Mn yields on metallicity.

As discussed in Section~2.1, we concentrate on the early enrichment
of the halo gas up to 1 Gyr of evolution, where the dominant role is
played by SNII. We adopt the SNII yields from WW95 for the mass range
12--40 \msb at $Z=10^{-4}Z_\odot$ and at $Z=0$ (three different
explosive treatment are presented in WW95:  as in Sect.~2.1, we use the
models labelled ``A'', representing the standard explosive treatment).
According to WW95, zero-metallicity models of 35 and 40 \msb provide
[Co/Mn] ratios similar to the higher metallicity models of
corresponding mass, whereas the 25 and 30 \msb moldels of
zero-metallicity produce higher ratios ($\sim$ 1 dex). For these
reasons, and given the difficulties in modelling the explosion of the
most massive stars, we assume that only stars with $m \simeq 40$~\msb
contribute to the enrichment of the proto-Galactic gas in Fe-group
elements.

The results obtained with our stochastic model for the evolution of Co
and Mn during the first Gyr of evolution of the Galaxy are shown in
Fig.~7, compared with spectroscopic observations of metal-poor stars
from McWilliam et al.~(1995b) and Ryan et al.~(1996) (stars observed by
both authors are connected by dotted lines). For the extremely
metal-poor star CS 22876-032 (with [Fe/H] $=-3.71$), recent spectroscopic
data by Norris, Beers, \& Ryan~(2000) indicate a metallicity higher
than the values estimated by Ryan et al.~(1996) because of the
decomposition of the spectrum into two components (the star has been
classified as double line spectroscopic binary), and a much lower Mn
abundance.

\subsection{Europium and Barium Enrichment in the Halo Gas}

The scatter observed in the relative abundances of neutron capture
elements with respect to iron, e.g. [Eu/Fe] and [Ba/Fe], is often
interpreted as an evidence for the inhomogeneous enrichment of the ISM
at the very beginning of the evolution of the Galaxy.  The abundance
distribution of [Eu/Fe] provides a direct way to address this problem,
since Eu is mostly produced by $r$-process nucleosynthesis. For Ba, the
contribution of $s$-process nucleosynthesis from AGB stars in the mass
range 2--4~\ms (hereafter low-mass stars, or LMS) accounts, at the
epoch of solar formation, for 80\% of the known abundance (for more
details see e.g. Travaglio et al.~1999, Gallino et al.~1998), whereas 
only $\sim 20$\% comes from the $r$-process.

The quantitative estimates for the $r$-process yields of Eu and Ba are
taken from Travaglio et al.~(1999). These authors explored different
SNe mass range for the production of $r$-process in the contest of a
chemical evolution model. They noticed that in order to reproduce the
typical increasing trend of [Ba/Fe] and [Eu/Fe] at [Fe/H] $\simeq
-1.5$, the production of Fe at low metallicity must have occurred
mainly before the production of $r$-process component of Ba and Eu.
According to their model, SNII in the mass range 8 -- 10 $M_\odot$
appear to be good candidates for the primary production of $r$-nuclei,
whereas a range extending to much higher masses seems to be in conflict
with the available observations.  As noticed by the authors, the
assumption of $r$-process from low-mass SNe is also supported by recent
theoretical models by Wheeler et al.~(1998) and Freiburghaus et
al.~(1999).

For this work we adopt the same assumptions on the $r$-process yields
from SNII in the mass range 8--10~\msb, as listed in Table~2. In
Fig.~8 we show the results of the Monte Carlo model for the Eu
enrichment in the halo gas from 8--10~\msb at different metallicities.
This Figure provides more insight into the age at which the halo gas
begins to be enriched by the $r$-process, as well as the metallicity
range that star forming clouds enriched in Eu can cover at a given
epoch. In fact one can see that the first enrichment of Eu in the
clouds starts at [Fe/H] $\simeq -5$, corresponding to $\sim 10^8$~yr.
This delay is due to the choice of the SNe mass range for the
production of $r$-process.  At [Fe/H] $\simeq -2$, the Eu composition
is mostly homogenized (for the last $10^8$ yrs). In this Figure is also
shown for comparison (with a solid line) the chemical enrichment of Eu
obtained with the Travaglio et al.~(1999) model. One can see that the
higher $X$(Eu) values obtained with the Monte Carlo model at 
$10^8$--$10^9$~yr are well matched by the average $X$(Eu) value obtained with
the Travaglio et al.~(1999) model.  However with the Monte Carlo model
the gas is enriched in Eu since lower metallicities. In fact with the
Monte Carlo model we can simulate clouds enriched in Eu at [Fe/H]
$\simeq - 5$ but formed at later times with respect two clouds with
higher [Fe/H].

The Eu enrichment in the halo gas is also shown in Fig.~9, together
with the Ba enrichment, in terms of [Eu/Fe] and [Ba/Fe] vs.  [Fe/H],
compared with spectroscopic observations. The observational sensitivity
limit for Eu and Ba is given by [Eu/H] $\simeq -4\div -3$ and [Ba/H]
$\sim -6\div -5$ (Gratton, private communication), and is shown by a
long-dashed lines. At [Fe/H] $\simeq -2$ the Eu and Ba enrichment of
the gas has dropped (i.e. the contribution of the 8--10~\msb has
vanished), and the Fe composition started to be dominant.  This results
in a decrease of [Eu/Fe] and [Ba/Fe] at [Fe/H] $\ge -2$, and a
contribution from the thick disk is required. At [Fe/H] $\simeq -2$ an
additional (and dominant) contribution to Ba from $s$-process
nucleosynthesis in low-mass AGB stars is also required (as shown by
Gallino et al.~1998 and Travaglio et al.~1999, the complex behavior
of the $s$-element yields as a function of metallicity allows them to
dominate the Galactic enrichment only at [Fe/H] $\ge -2$).

As discussed in the previous Section and shown in Fig.~6, a higher
SF efficiency (e.g. 30\% of the mass of the clouds converted in stars)
will increase the final [Fe/H] at 1 Gyr up to $\sim 1$~dex. The effect
on [Eu/Fe] and [Ba/Fe] is shown in Fig.~10, where one can see that the
Monte Carlo model predicts a spread in [Eu/Fe] and [Ba/Fe]
composition where the observations show a more homogeneous composition.

Following the discussion presented by Travaglio et al.~(1999), we have
also explored different SNII mass ranges for the production of
$r$-process elements. We show in Fig.~11 the result obtained under
the assumption that $r$-process nucleosynthesis occurs in 15--30~\msb
SNII.  In this case the enrichment of the gas in Eu and Ba starts at
earlier times with respect to the simulation shown in Fig.~8, due to
the shorter lifetimes of this SN mass range. However, the fraction of
stars in this mass range, as well as the $r$-process yields, produce an
average value of [Eu/Fe] and [Ba/Fe] vs. [Fe/H] which is higher than
the case with a low SNII mass range. This result is in disagreement
with the observed values of [Eu/Fe] and [Ba/Fe]. The only exception is
represented by the peculiar star CS 22892-052 (where $r$-process
elements are enhanced over about 40 times the solar value), characterized by
[Fe/H] $\simeq -3$, [Eu/Fe] $\simeq +1.44$, and [Ba/Fe] $\simeq
+0.93$.  With the $r$-process yields adopted here these values are in
general not reproduced, unless the $r$-process come from higher mass
range SNII (as shown in Fig.~11), at the price of not reproducing the
abundances observed in the other stars.

We have also explored the consequences of zero Fe yields for higher
mass stars.  From the point of view of stellar evolution, this is
supported by the idea that mass-loss increases with metallicity (the
efficiency of radiation pressure driven wind scales with metallicity as
$\dot{M}\propto Z^{0.5}$, see Kudritzki et al.~1989) and that the
probability to have even less Wolf-Rayet stars that produce Fe is
higher at lower metallicities. However, this will not affect
considerably the results for the inhomogeneous composition of the gas
at $t \geq 10^8$ yr, when Eu production starts. This is due to the
short lifetimes of these massive stars which lead to chemical
enrichment of the ISM on a shorter timescale ($\simeq$ few times
$10^6$~yr).

\subsection{Strontium Enrichment in the Halo Gas}

The study of the chemical evolution of Sr deserves a deeper analysis
and special comments. 
A large number of metal-poor stars with [Fe/H] $\le -2.0$ and $-1.5
\le$ [Sr/Fe] $\le +0.8$, shows a relative abundance of Sr that tends to
decrease with an increase of iron abundance (see e.g.  Ryan~2000, with
references therein for a collection of data, and Burris et al.~2000).
However, the estimate of the different sources that contribute to the
Galactic evolution of Sr has not yet quantified in detail.

Gallino et al.~(1999) analyzed the various contributions to the solar
abundance of $s$-process elements, arguing that while low-mass AGB
stars can account almost entirely for the heavy $s$-isotopes from Ba to
Pb, they can contribute only $\sim$ 70\% of the solar composition of
nuclei around Sr.  The addition of $s$-process yields from intermediate
mass AGB stars (with $m > 4$~\ms, hereafter IMS) can play a relevant
role in the Galactic nucleosynthesis of the elements across the Sr-Y-Zr
peak and provide up to 15\% of their solar system abundances (Gallino
et al.~1999). We should however keep in mind that the mechanisms of
production of $s$-process nuclei from IMS are not not completely 
elucidated and that detailed models have become available only very
recently.

In the halo phase the effects of the AGB nucleosynthesis on the
chemical enrichment of the ISM are less important with respect to the
disk, due to the relatively long time scale required by their
evolution.  In the halo the production of heavy elements is mostly due
to the $r$-process (see e.g. Truran~1981).  High-mass stars also
contribute to the Sr enrichment of ISM, via $r$-process nucleosynthesis
(see e.g.  Woosley et al.~1994), in addition to a small contribution by
$s$-process from massive stars ($\sim 5$\%, according to Raiteri et
al.~1993). The fact that the $r$-process is not as simple as currently
assumed and requires different sites for producing nuclei respectively
lighter and heavier that $A \simeq 140$ was first advanced by
Wasserburg, Busso \& Gallino~(1996), but presnt-day stellar models are
not yet able to quantify the production of these elements. Looking at
the observations of Sr, the lower boundary of the observed [Sr/Fe]
abundance can be explained by the ``normal'' $r$-process behaviour,
corresponding to a universal [Ba/Sr] $r$-process value.  An additional
source is required to explain the large spread observed in [Sr/Fe], as
for example a primary component from massive stars. The observations of
Sr strongly suggest that the process at work must involve low neutron
exposures that synthesize only species around the atomic number of Sr,
without reaching species with atomic number around Ba. The observations
of other species near Sr, also enhanced in these high Sr stars (Ryan et
al.~1996), provide additional evidence supporting this possibility.

To explore the spread in [Sr/Fe] vs. [Fe/H] that we can reproduce with
the Monte Carlo model presented here, we summarize the contribution to
Sr from different stellar sources during the evolution of the Galaxy,
as discussed above:

({\em i}\/) an $s$-component from LMS
($m \simeq 2$--4~\ms), for $\sim 60$\% of the the solar composition;   

({\em ii}\/) an $s$-component from IMS
($m \simeq 4$--7~\ms), for $\sim 15$\% of the solar composition;

({\em iii}\/) an $s$-component from massive stars, called {\it
weak-component},
for $\sim 5$\% of the solar composition;

({\em iv}\/) a primary component to be attributed to massive stars
with $m \simeq 15$--25~\ms, for $\sim$ 20\% at any time.

Notice that since ({\em iii}\/) is a secondary component, i.e. it needs
an enrichment of the gas from a previous generations of stars, it is   
negligible at low metallicities.

Fig.~12 shows the result for [Sr/Fe] vs. [Fe/H] computed under these
assumptions. Since we are interested in the early phases of evolution
of the Galaxy we took into account only the Sr contribution from
massive stars.  The spread observed in low metallicity stars is well
reproduced by the model up to [Fe/H] $\simeq -1.8$. As in Fig.~9 and
Fig.~10 we plotted an observational limit for Sr given by [Sr/H] in
the range $\simeq -6\div -5$ (Gratton, private communication).

\section{Conclusions}

Spectroscopic abundance determinations in various samples of metal-poor
halo stars reveal the presence of large dispersions in heavy elements. If
the observed scatter is {\it intrinsic} (not substantially influenced
by different calibration methods or data reduction procedures), this
fact can be interpreted as a signature of incomplete mixing of the
interstellar medium at the first epochs of the Galaxy.

In this work we have presented a chemo-dynamical model, based on a
Monte Carlo technique, specifically devoted to study the inhomogeneous
evolution of the Galactic halo. We have considered the effects of local
enrichment of the halo gas by star formation episodes in interstellar
clouds, which subsequently {\it fragment} into smaller objects, as well
as the {\it coalescence} between clouds.  With this model we explored
the SFR versus time, the interstellar mass function, the
age-metallicity relation and the G-dwarf distribution.  We have also
discussed the effects of a term of exponential infall (with an
e-folding time of 10$^9$ yr) from the halo gas to the disk, and the
evolution of the mass spectrum of coalesced interstellar clouds.

The main parameters that control the dynamical evolution of the halo
as well as the enrichment history of the halo ISM are the initial mass
range of clouds (from 10$^3$ \msb up to 10$^7$ \msb, chosen in order to
obtain a SFR $\sim 1$~\ms/yr), the efficiency of star formation (a few
percent of the mass of the clouds), and the frequency of coalescence
episodes with respect to star formation (the standard choice for this
work is $f_{\rm m}=2f_{\rm b}$).

The main goal of the model is to follow the evolution of the halo
metallicity and of the heavy elements Eu, Ba and Sr during the early
ages of the Galaxy.  Since this work is concentrated on the halo
evolution, we have quantified the $r$-process contribution, using
previous analytical calculations to derive the $r$-process yields
introduced by Travaglio et al.~(1999).  The comparison of the model
results with available spectroscopic data for Population II stars
suggests a production of Eu and Ba from SNII, slightly delayed with
respect to the main phase of oxygen enrichment, that can be explained
in the context of low-mass (8--10~\ms) SNII.

Under these assumptions, the values of [Eu/Fe] and [Ba/Fe] cover a
spread of $\sim$ 3 dex up to [Fe/H] $\simeq -2$. At higher [Fe/H]
(corresponding to $t \ge 4\times 10^8$~yr), the gas has had enough time
to homogenize its chemical composition and the large [element/Fe]
spread is considerably reduced. Two interesting points should be
noticed:  first, the model predicts the existence of a large number of
clouds with [Eu/Fe] and [Ba/Fe] values 1 or 2 dex lower than the
observations, below the present-day threshold for spectroscopic observations of Eu
and Ba (Gratton, private communication). Moreover, the model predicts
halo gas with metal-enrichment of [Fe/H] $< -4$, but no stars are
observed at such low metallicities. One possible explanation, is again
the difficulty to observe objects with such a low metallicity; another
possibility is a pre-enrichment of the halo gas by a Population III of
massive stars.

We have also discussed the problem of chemical evolution of Sr. The
production of this element was obtained under different assumptions
with respect to Ba and Eu. Our explanation of the spread of [Sr/Fe]
observed in low metallicity halo stars is that in addition to a
$s$-contribution from LMS and IMS, and a $s$-contribution from massive
stars (weak-component), there is a primary component from SNII with
masses $\sim$ 15--25~\msb accounting for $\sim 20$\% of the Sr solar
composition.

\acknowledgements
The research of C.T. and D.G has been supported in part by grant
COFIN98-MURST at the Osservatorio di Arcetri.  C.T. would like to thank
R. Gratton for kind comments about the spectroscopic observations, and
A. Nelson for useful discussions at the beginning of this work.

\newpage

\begin{figure}
\plotone{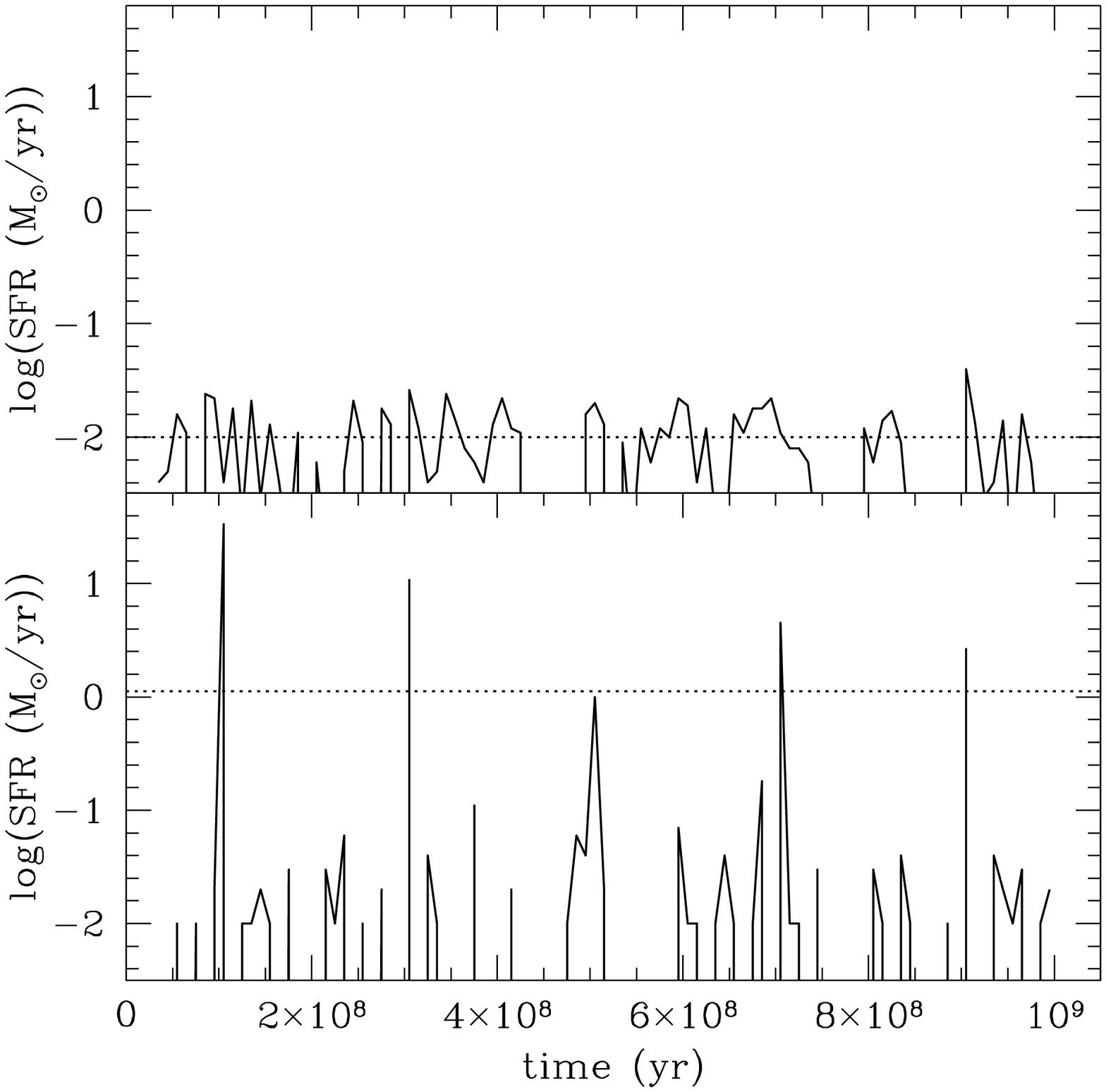}
\figcaption[fig1.eps]{Star formation rate (in \ms~yr$^{-1}$) vs. time
for two different assumptions on the initial mass of the clouds in the
range $10^3$--$10^5$~\ms ({\it upper panel}), and $10^3$--$10^7$~\ms
({\it lower panel}).  The {\it short-dashed} line represents the
average value.}
\end{figure}

\newpage

\begin{figure}
\plotone{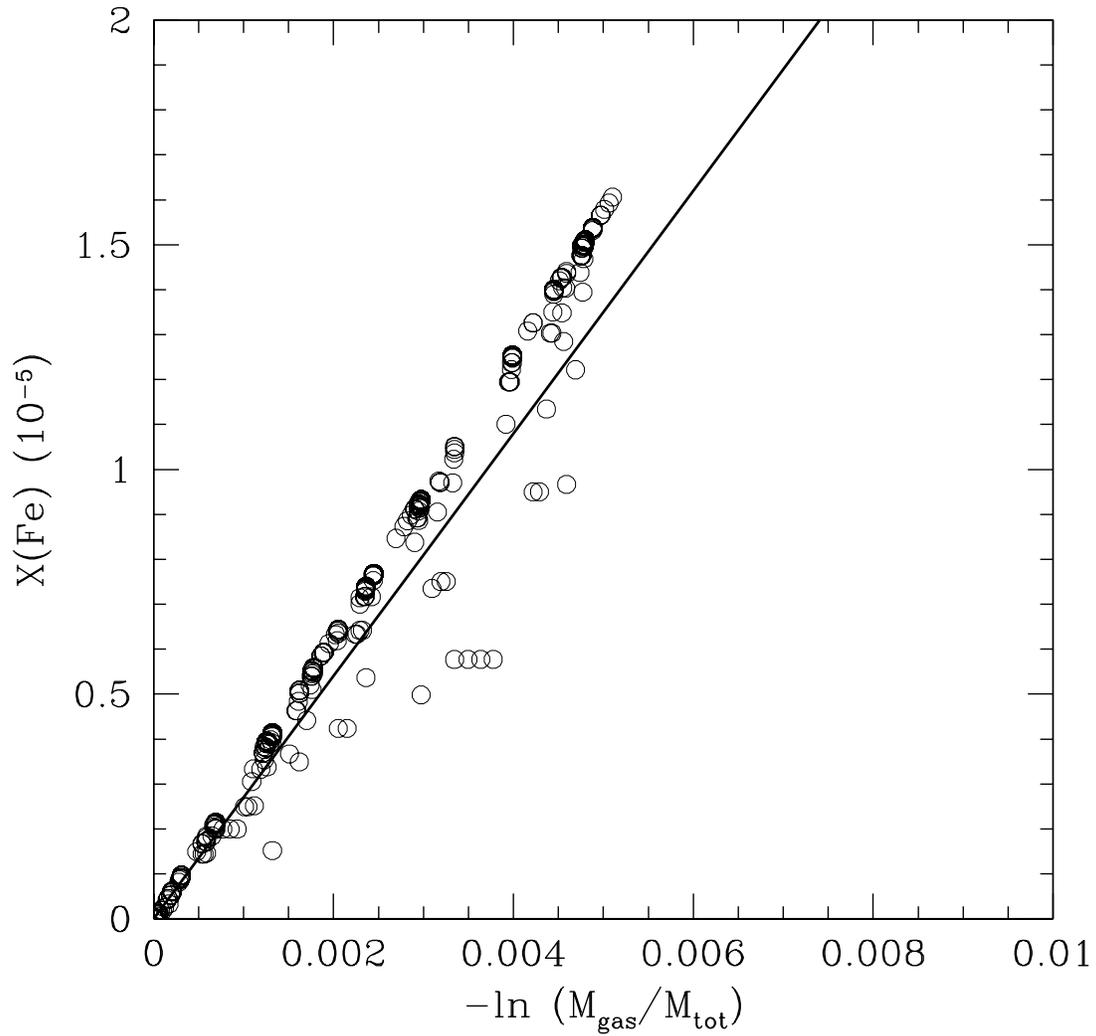}
\figcaption[fig2.eps]{Abundance of Fe as function of the gas fraction
for the Monte Carlo model without infall ({\it open circles}) compared
with the simple closed box model in the IRA ({\it continuous line}).
For the Monte Carlo model $X$(Fe) is computed as the sum of the Fe mass
present in all the clouds divided by the total mass of the gas at each
timestep.}
\end{figure}

\newpage

\begin{figure}
\plotone{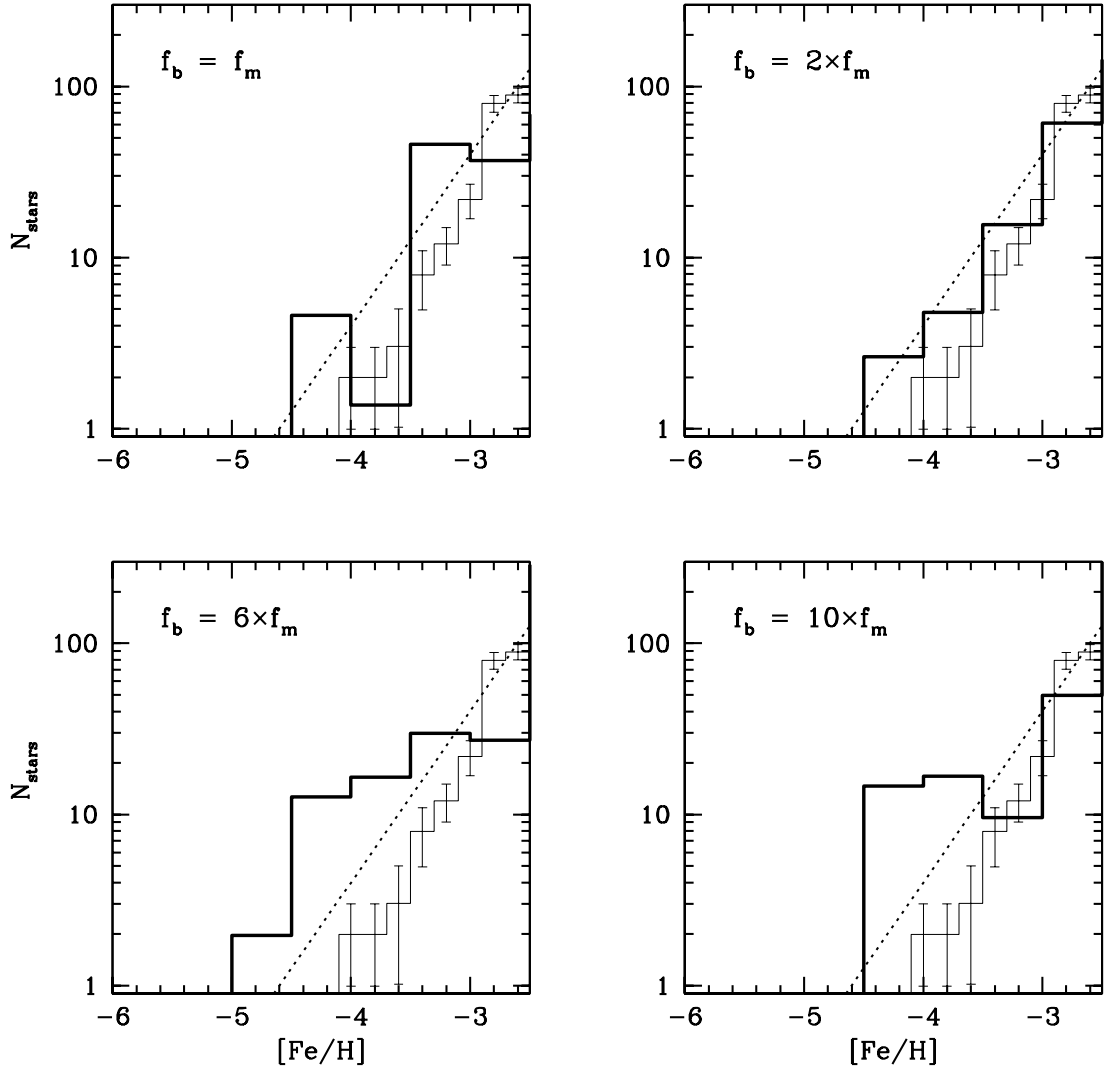}
\figcaption[fig3.eps]{G-dwarf distribution for halo stars. The result
of the Monte Carlo model ({\it solid line}) is compared with the
collected sample of observations by Beers et al.~(1998) ({\it dotted
line}), and to the simple model of chemical evolution in the IRA
approximation ({\it dashed line}). The errorbars for the observational
data indicate the $\sqrt{N_{stars}}$ noise associated with each bin.
The number of G-dwarf stars is plotted in arbitrary units. The four panels
show four different choices for the ratio between the frequency of
coalescence $f_{\rm m}$ and of star formation $f_{\rm b}$, as discussed in
the text.}
\end{figure}

\newpage

\begin{figure}
\plotone{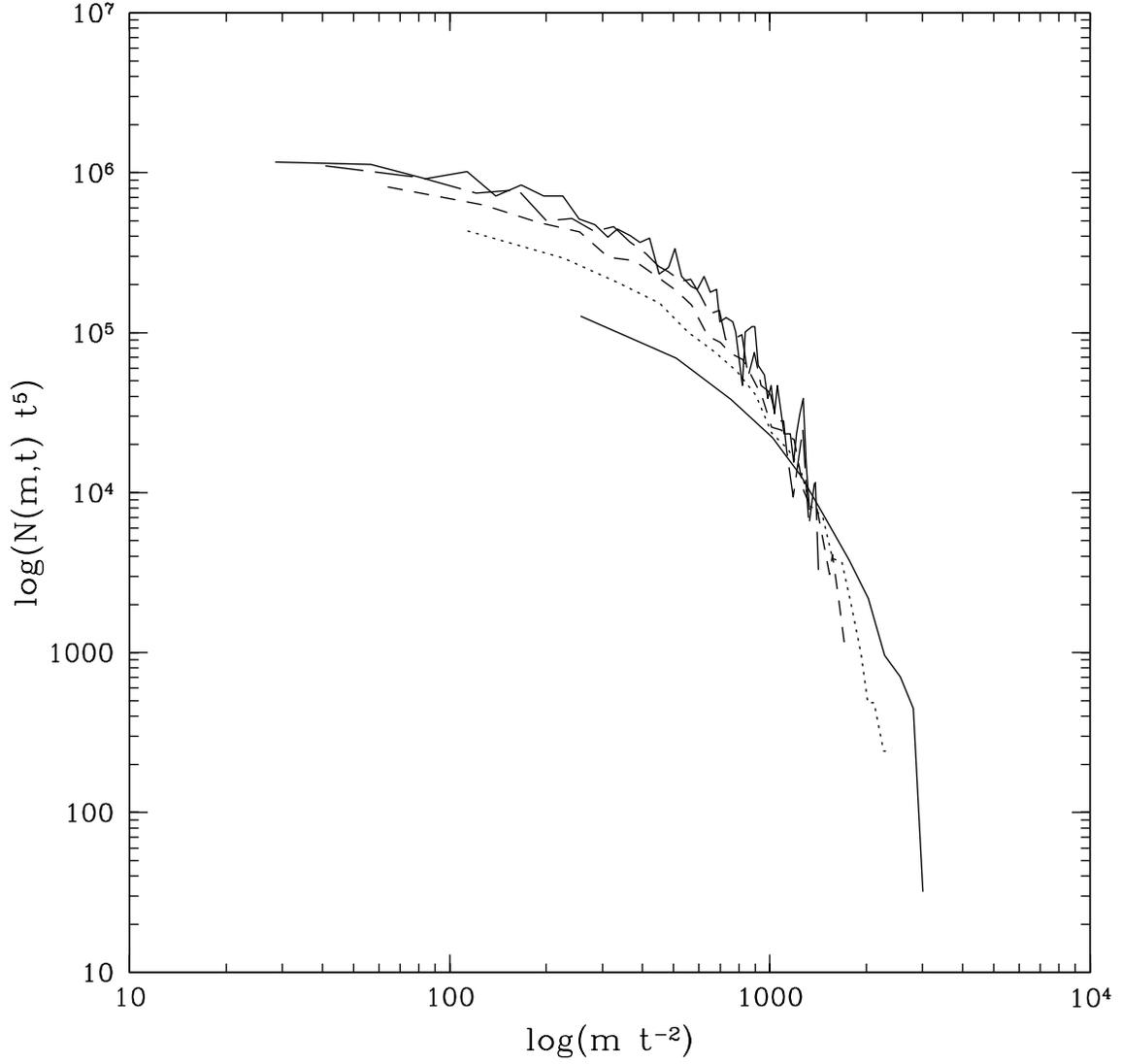}
\figcaption[fig4.eps]{Evolution vs. time of the mass spectrum of
coalesced interstellar clouds in a similarity solution according to eq.
(19) in the text and for $\lambda=1$. Different lines correspond to
different timestep of evolution.}
\end{figure}

\newpage

\begin{figure}
\plotone{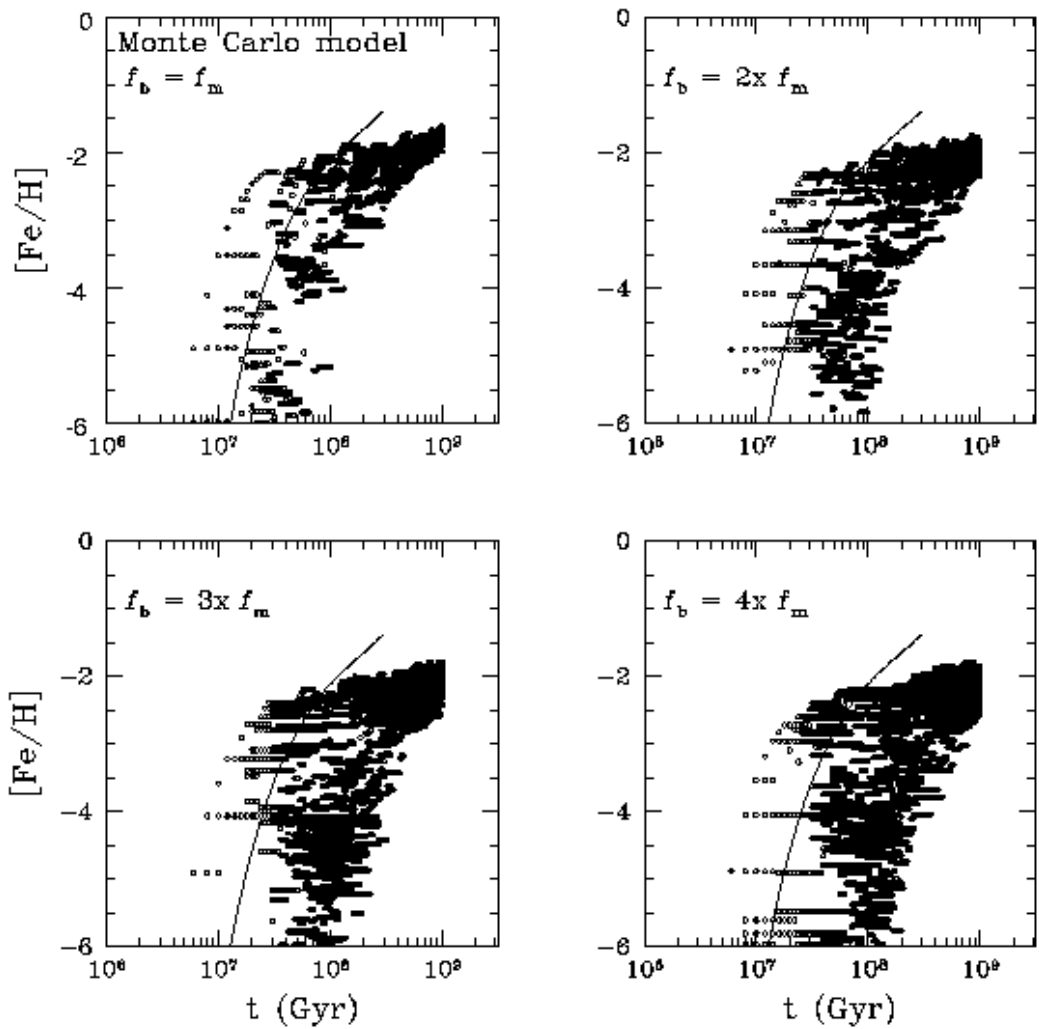}
\figcaption[fig5.eps]{Distribution of metallicity vs. time obtained
with the Monte Carlo model. The four panels show four different choices
for the ratio between the frequency of coalescence $f_{\rm m}$ and of
star formation $f_{\rm b}$, as discussed in the text.  Points represent
interstellar clouds during the first Gyr of the evolution of the
Galaxy. Solid line represents the metallicity evolution obtained by Travaglio et 
al.~(1999).}
\end{figure}

\newpage

\begin{figure}
\plotone{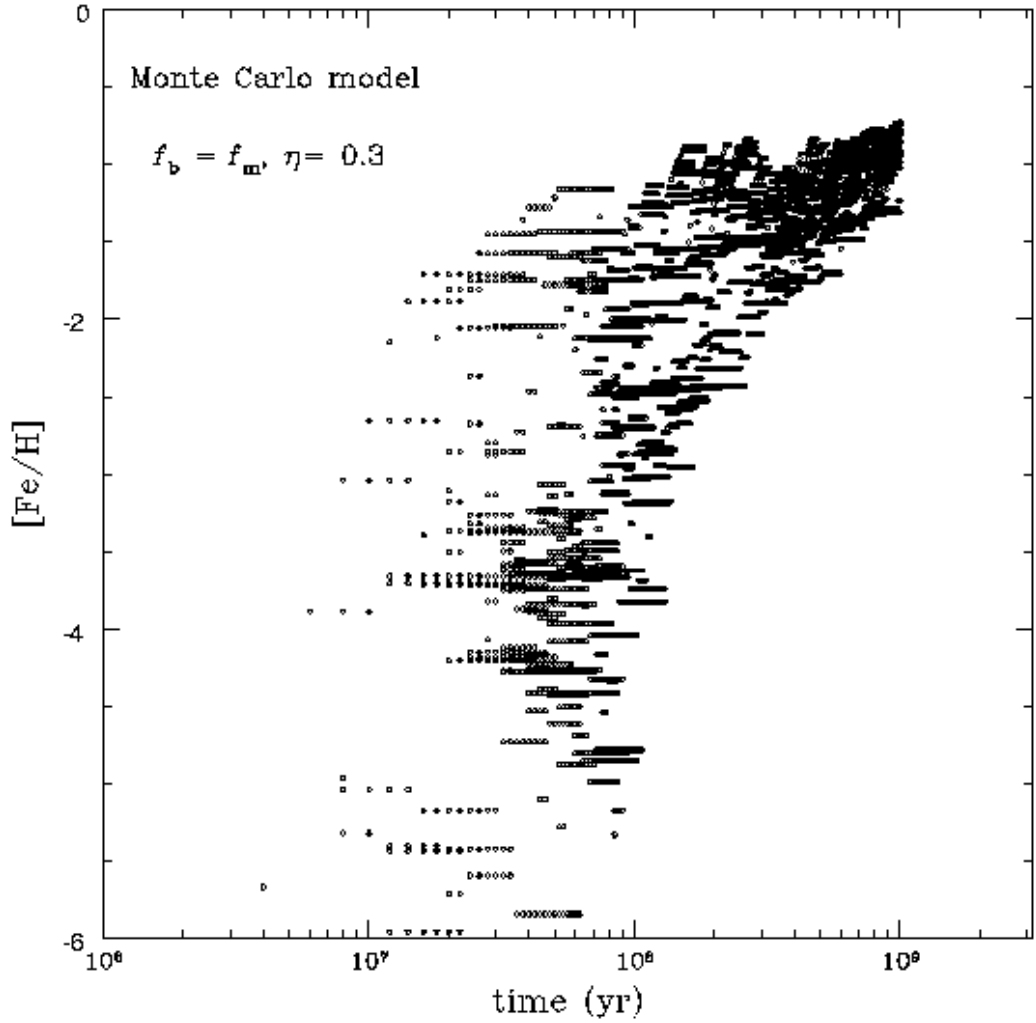}
\figcaption[fig6.eps]{Distribution of metallicity vs. time obtained
with the Monte Carlo model for the choice of frequency of coalescence
$f_{\rm m}$ equal to the frequency of star formation $f_{\rm b}$, and with
efficiency $\eta=0.3$}
\end{figure}

\newpage

\begin{figure}
\plotone{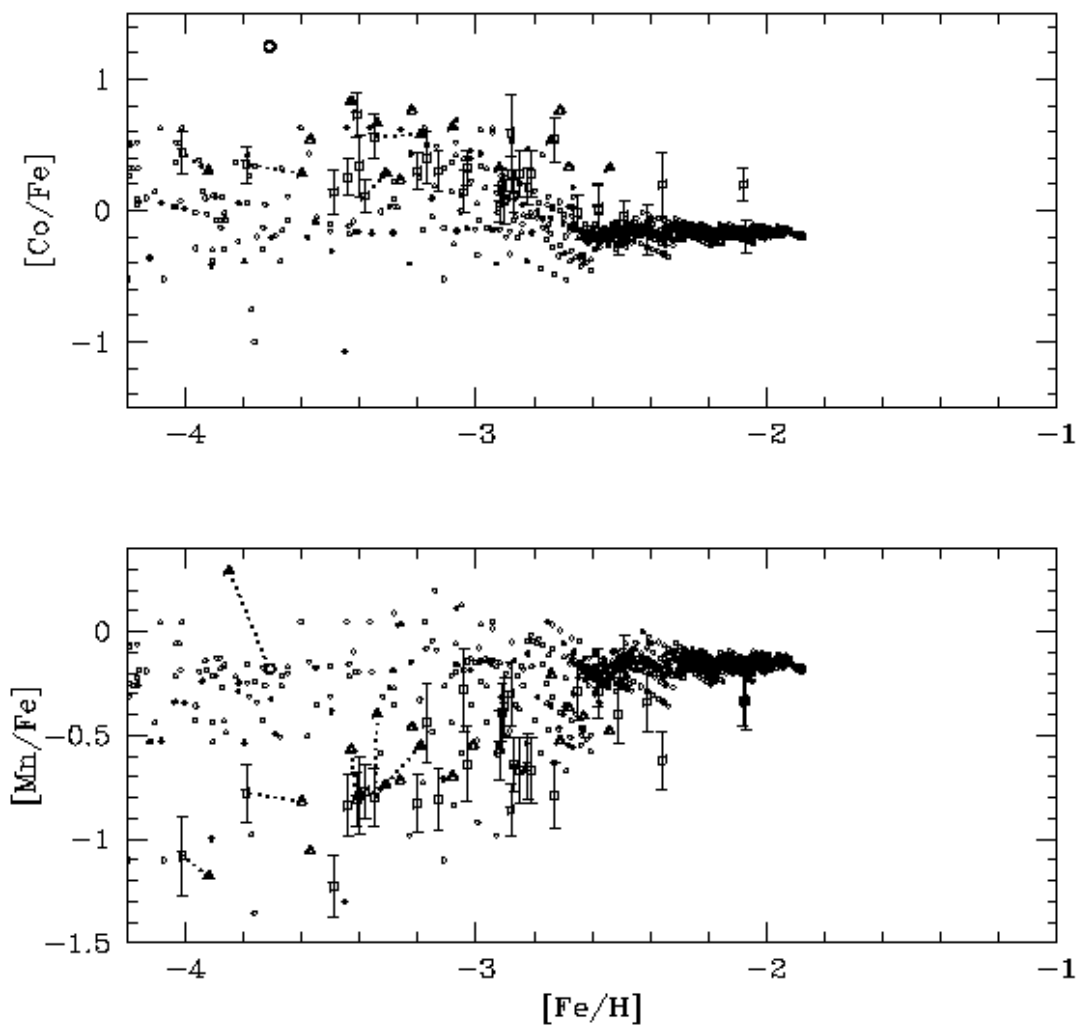}
\figcaption[fig7.eps]{Enrichment of Co ({\it upper panel}) and Mn ({\it
lower panel}) in the halo gas at different metallicities. Small
open circles represent interstellar clouds in the Monte Carlo model.
Observational data are from: McWilliam et al.~(1995b) ({\it open
squares}), Ryan et al.~(1996) ({\it open triangles}), and Norris et
al.~(2000) ({\it open circles}). Thin dotted lines connect same stars with
different abundance determinations. Errorbars are shown only when reported
by the authors for single objects.}
\end{figure}

\begin{figure}
\plotone{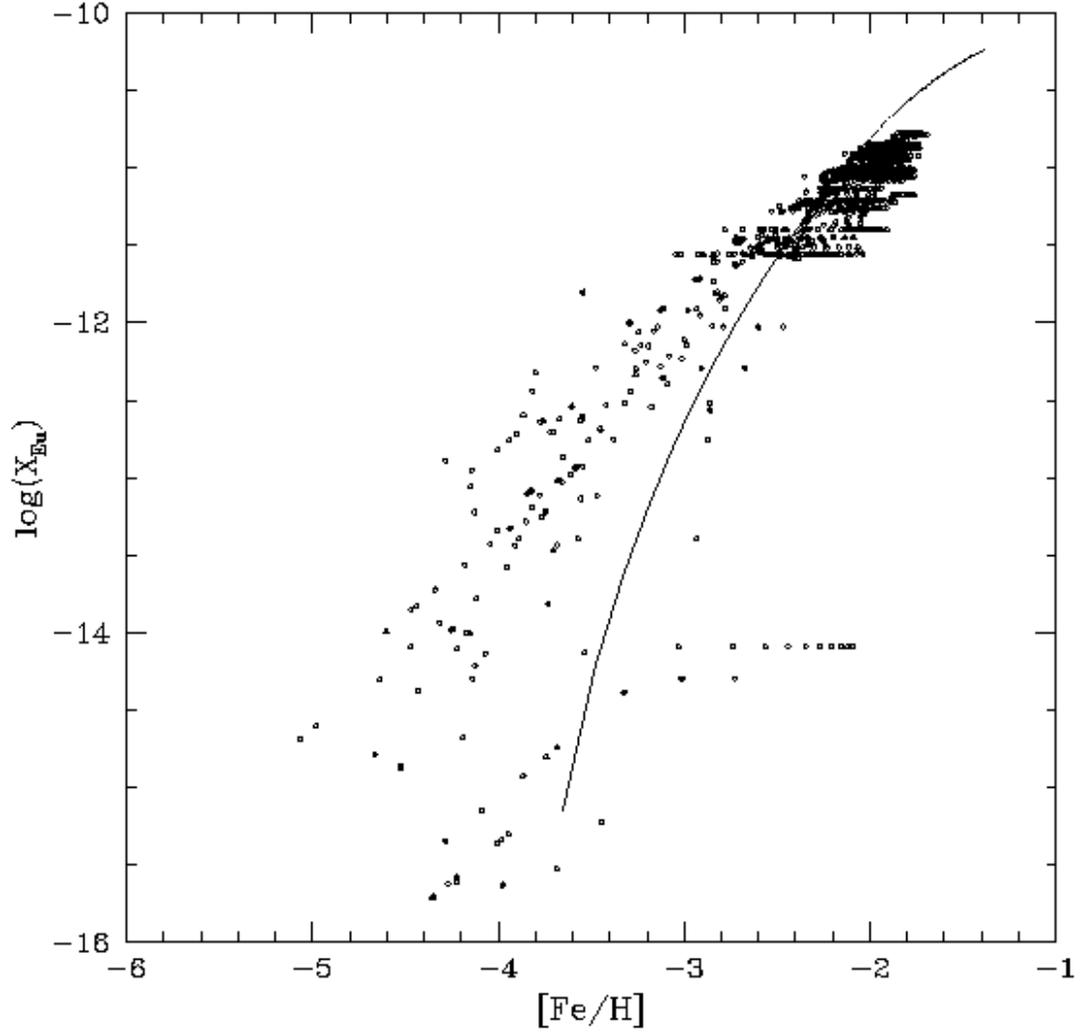}
\figcaption[fig8.eps]{Enrichment of Eu in the halo gas at different 
metallicities during the first Gyr of Galactic evolution. Small
open circles represent interstellar clouds in the Monte Carlo model.
The {\it solid line} represents the one-zone halo model presented 
in Travaglio et al.~(1999).}
\end{figure}

\begin{figure}
\plotone{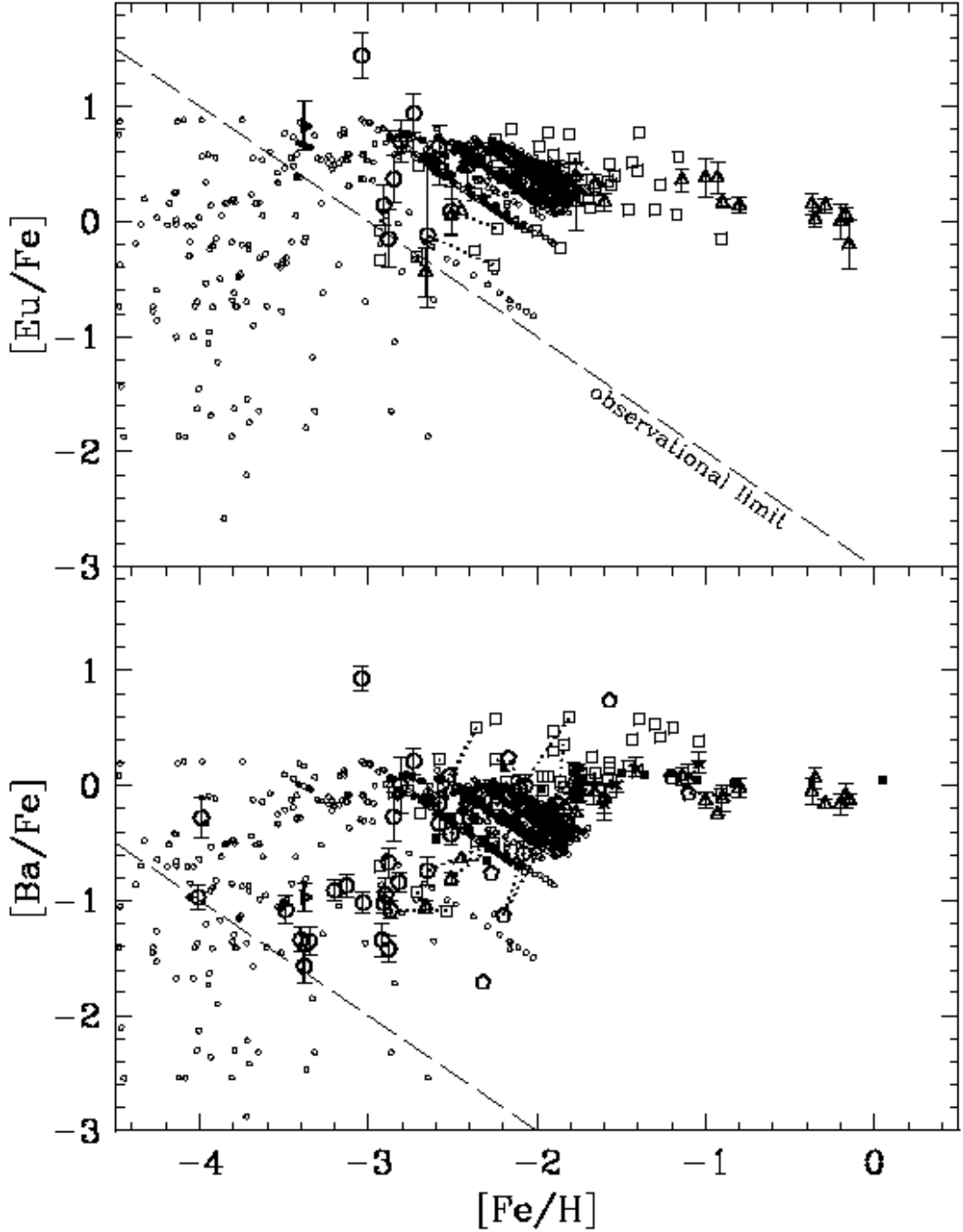}
\figcaption[fig9.eps]{{\tiny Abundance of Eu ({\it upper panel}) and Ba
({\it
lower panel}), as a function of [Fe/H], during the first Gyr of
evolution of the Galaxy (model predictions are shown as {\it small thin
open circles} and represent our best-fit model with the choice
parameters reported in Table~1).  The {\it long-dashed line} indicates
the sensitivity limit for observing Eu and Ba. The $r$-process yields
of Eu are derived from SNII in the mass range 8--10~\ms. Observational
data of Eu and Ba in metal-poor stars are from: Gratton \&
Sneden~(1994) ({\it open triangles}); Woolf et al.~(1995) ({\em open
squares}); Fran\c cois~(1996) ({\em pentagons}); McWilliam et
al.~(1995) and McWilliam~(1998) ({\em thick open circles}); Norris,
Ryan \& Beers~(1997) ({\it open tilted triangles}); Jehin et al.~(1998)
({\em filled tilted triangles}); Mashonkina et al.~1999 ({\em filled
squares}); Burris et al.~(2000) ({\em open squares}). Thin dotted lines
connect stars with different abundance determinations.  Errorbars are
shown only when reported by the authors for single objects.}}
\end{figure}

\newpage

\begin{figure}
\plotone{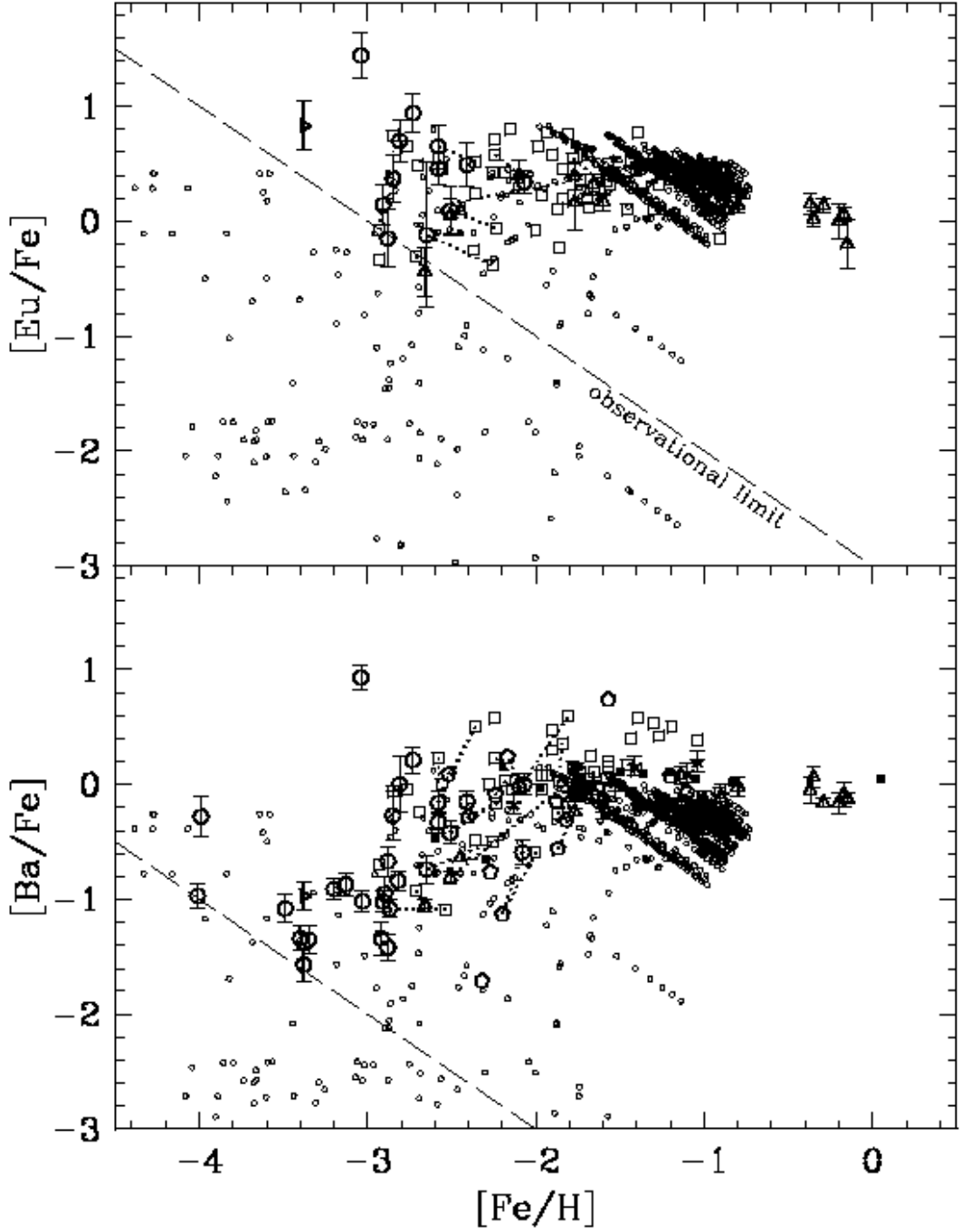}
\figcaption[fig10.eps]{The same of Fig.~10, for star formation 
efficiency $\eta=0.3$ (see text for discussion)}.
\end{figure}

\newpage

\begin{figure}
\plotone{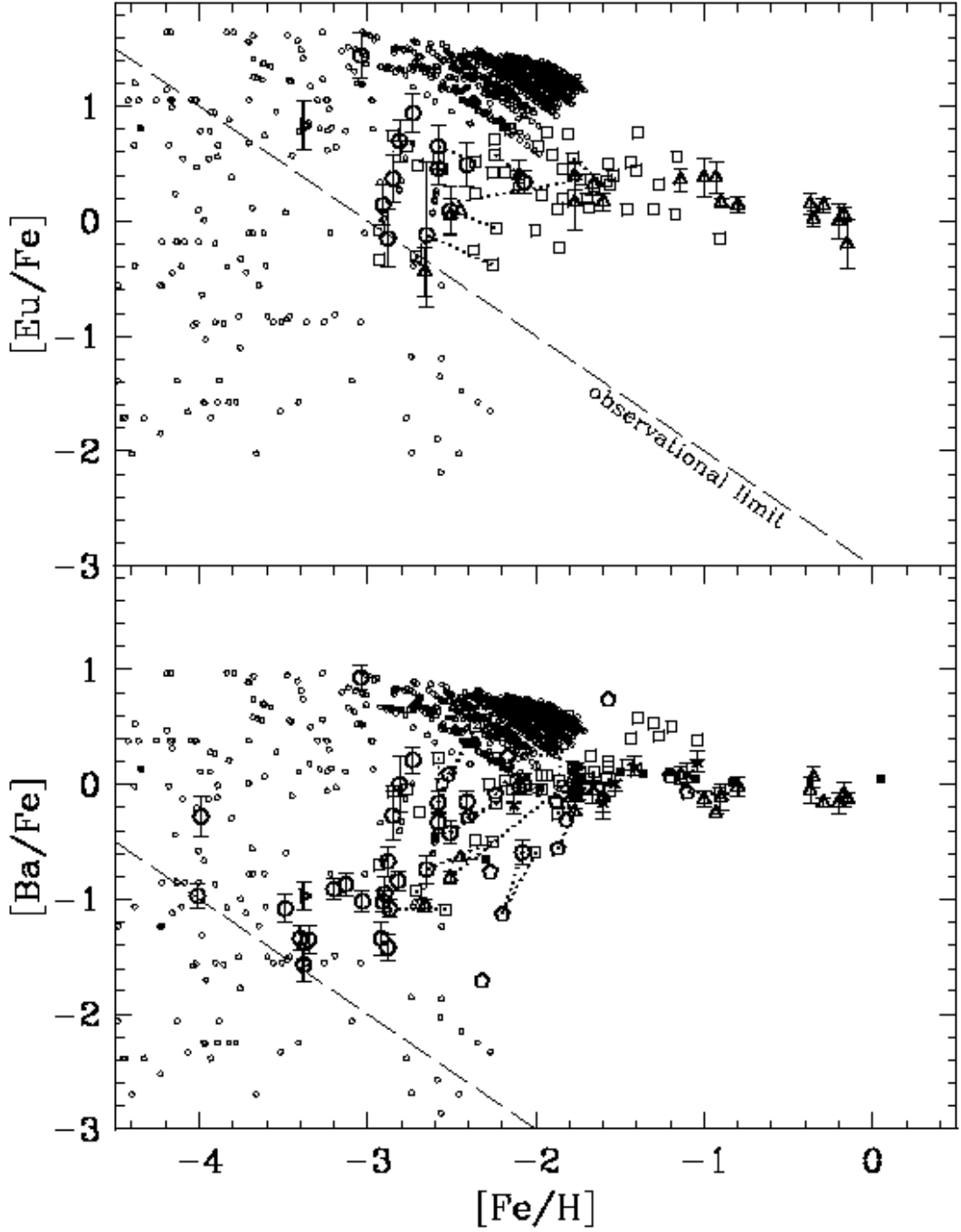}
\figcaption[fig11.eps]{The same of Fig.~9, in which Eu and Ba are
produced by SNII in the mass range 15--30~\ms.}
\end{figure}

\newpage

\begin{figure}
\plotone{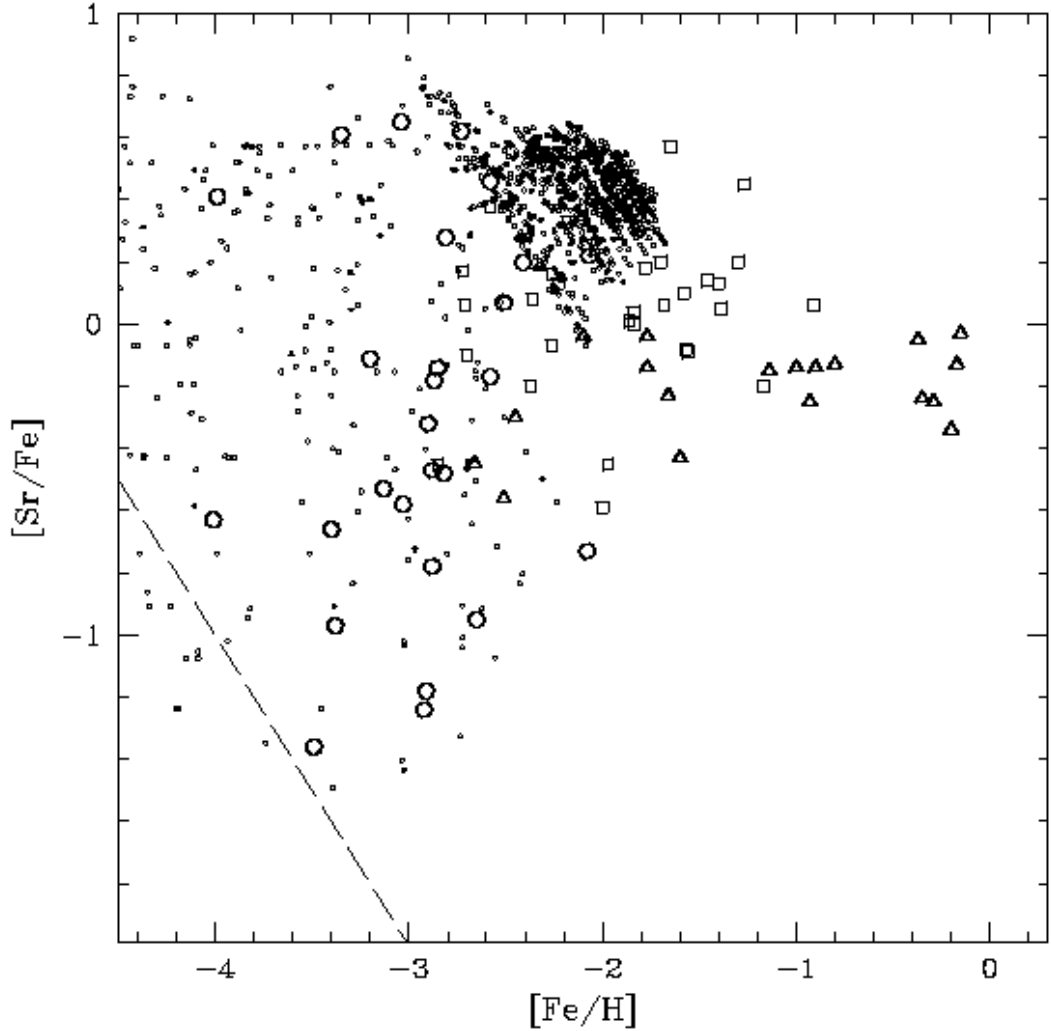}
\figcaption[fig12.eps]{Chemical evolution of Sr, under the assumption
described in the text, and with an additional primary component from
SNII with masses $\sim 15$--25~\ms, accounting for $\sim 20$\% of the
solar Sr abundance. The Monte Carlo model results are shown as {\it
small thin open circles}. Observational data are from: Burris et
al.~(2000) ({\em open squares}); McWilliam~(1998) ({\em open circles});
Gratton \& Sneden~(1994) ({\em open triangles}). The {\it long-dashed
line} indicates the sensitivity limit for observing Sr.}
\end{figure}

\end{document}